\documentclass[useAMS,usenatbib]{mn2e}
\RequirePackage{fixltx2e}
\usepackage[a4paper,top=1in,bottom=1in,left=15mm,right=15mm,footskip=0.25in]{geometry}
\usepackage[backref,breaklinks,colorlinks,linkcolor=blue,citecolor=blue]{hyperref}      

\usepackage[all]{hypcap}

\providecommand{\adsurl}[1]{\href{#1}{ADS}}
 
\providecommand{\url}[1]{\href{#1}{#1}}
\usepackage{url}

\usepackage{amsmath}
\usepackage[usenames,dvipsnames]{color}
\usepackage{graphicx}
\usepackage{amsmath}
\usepackage{float}

\usepackage{times}
\usepackage{mathptmx}

\providecommand{\abs}[1]{\lvert#1\rvert}
\providecommand{\E}{E}
\providecommand{\Eg}{E_{\rm g}}

\def\msun{{\,M_\odot}}

\def\alt{\raise0.3ex\hbox{$\;<$\kern-0.75em\raise-1.1ex\hbox{$\sim\;$}}}
\def\agt{\raise0.3ex\hbox{$\;>$\kern-0.75em\raise-1.1ex\hbox{$\sim\;$}}}

\newcommand{\bw}{\begin{widetext}}
\newcommand{\ew}{\end{widetext}}

\newcommand{\lsim}{\,\rlap{\raise 0.35ex\hbox{$<$}}{\lower 0.7ex\hbox{$\sim$}}\,}
\newcommand{\gsim}{\,\rlap{\raise 0.35ex\hbox{$>$}}{\lower 0.7ex\hbox{$\sim$}}\,}

\def\lesssim{\mathrel{\hbox{\rlap{\hbox{\lower3pt\hbox{$\sim$}}}\hbox{\raise2pt\hbox{$<$}}}}}
\def\gtrsim{\mathrel{\hbox{\rlap{\hbox{\lower3pt\hbox{$\sim$}}}\hbox{\raise2pt\hbox{$>$}}}}}

\def\xlinkspace#1 #2{%
 \ifx\relax#2%
 \xlinkdash#1-\relax
 \else
 \xlinkdash#1 -\relax
 \expandafter\xlinkspace\expandafter#2%
 \fi}

\def\xlinkdash#1-#2{%
 \ifx\relax#2%
 \tmp{#1}%
 \else
 \tmp{#1-}%
 \expandafter\xlinkdash\expandafter#2%
 \fi}

\title[Cosmic-Ray Models of the Ridge-Like Excess of Gamma Rays in the Galactic Center]{Cosmic-Ray Models of the Ridge-Like Excess of Gamma Rays in the Galactic Center}

\author[O. Macias, C. Gordon, R.M. Crocker and S. Profumo]{Oscar Macias$^{1}$\thanks{E-mail:oam13@uclive.ac.nz}, Chris Gordon$^{1}$, Roland M.~Crocker$^{2,3}$, and Stefano Profumo$^{4,5}$\\
$^1$ Department of Physics and Astronomy, Rutherford Building, University of Canterbury,\\ Private Bag 4800, Christchurch 8140, New Zealand\\
$^2$ Research School of Astronomy and Astrophysics, Australian National University, Canberra, Australia\\
$^3$ Future Fellow\\
$^4$ Department of Physics, University of California, Santa Cruz, 1156 High Street, Santa Cruz, CA, 95064\\
$^5$ Santa Cruz Institute for Particle Physics, University of California, Santa Cruz, 1156 High Street, Santa Cruz, CA, 95064}

\begin{document}

\pagerange{\pageref{firstpage}--\pageref{lastpage}} \pubyear{2014}

\maketitle

\label{firstpage}

\begin{abstract}
The High-Energy Stereoscopic System (HESS) has detected diffuse TeV emission
correlated with the distribution of molecular gas along the Ridge at the Galactic Center.
Diffuse, non-thermal emission is also seen by the Fermi large area  telescope (Fermi-LAT) in the GeV range and
 by radio telescopes in the GHz range. Additionally, there is a distinct, spherically symmetric excess of gamma rays seen by Fermi-LAT in the GeV range. 
 A  cosmic ray flare, occurring in the Galactic Center,  $10^4$ years ago has been proposed to explain the TeV Ridge \citep{Aharonian:2006au}. 
 An alternative, steady-state model explaining all three data sets (TeV, GeV, and radio) invokes purely leptonic processes \citep{Yusef-Zadeh2013}. We show that the flare model from the Galactic Center also provides an acceptable fit to the GeV and radio data, provided the diffusion coefficient is energy independent. However, if Kolmogorov-type turbulence is assumed for the diffusion coefficient, we find that two flares are needed, one for the TeV data (occurring approximately $ 10^4 $ years ago) and an older one for the GeV data (approximately  $10^5$ years old).  
We find that
 the flare models we investigate do not fit the spherically symmetric GeV excess
 as well as the usual generalized Navarro-Frenk-White spatial profile, but are better suited to explaining
  the Ridge.
 {\color{black}	We also show that a range of single-zone, steady-state models are able to explain all three spectral data sets.
 	 Large gas densities equal to the volumetric average in the region 
	 can be accommodated by an energy independent diffusion} {\color{black} or streaming} {\color{black}	based steady-state model.}
 Additionally, we investigate how the flare and steady-state models may be distinguished with future gamma-ray data looking for a spatial dependence of the gamma-ray spectral index.
\end{abstract}


\section{Introduction}
\label{sec:introduction}

The High Energy Stereoscopic System (HESS) collaboration~\citep{Aharonian:2006au} reported the discovery of diffuse  TeV emission from the innermost part of the Galactic Center (GC) region. These gamma rays are localized over a ridge like area defined by Galactic longitude $\abs{ l } < 0.8^{\circ}$ and latitude $\abs{ b}<0.3^{\circ}$. Morphological analysis of the data shows a close correlation between gamma-ray emission and molecular gas present in the region, known as the central molecular zone (CMZ). This is a strong indication that the gamma rays originate in cosmic-ray (CR) interactions with  interstellar matter (and thus have a volumetric emissivity proportional to the density of CRs and that of target material).


\cite{Aharonian:2006au}  proposed a model for the TeV Ridge emission in which a single flare injection of CR protons occurring near the super-massive black hole Sgr A$^{\star}$ or the young supernova remnant Sgr A East could plausibly explain the observed TeV flux density.

Steady-state models for the TeV Ridge have also been proposed \citep{Crocker2011b, Yusef-Zadeh2013, YoastHullGallagherZweibel2014}. In these models there is continual emission of  CRs into the CMZ with multiple sources distributed within the CMZ rather than a single emission event at the GC.

Several groups, using data from the Fermi large area  telescope (Fermi-LAT), have identified an
extended GeV source of gamma rays in the GC region
\citep{Vitale:2009hr,Hooper:2010mq,Morselli:2010ty,Boyarsky:2010dr,hooper,Abazajian:2012pn,hooperkelsoqueiroz2012,gordonmacias2013,MaciasGordon2014,Abazajian2014,Daylan:2014}. 
It has been confirmed that this GeV extended source is best fitted by the combination of a spherically-symmetric template {\color{Black} ( radius
 $\sim 1^\circ$) }
and a ridge-like template resembling the CMZ in the inner $\sim 200$ pc of the GC~\citep{MaciasGordon2014}. For the spherical component, the square of a generalized Navarro-Frenk-White (NFW) profile with inner slope of $1.2$ provides a good fit, and for the GeV Ridge component, a template based on the HESS residuals or 20-cm continuum radio emission map provides a satisfactory fit. The spectrum of each template is not significantly affected in the combined fit and are consistent with previous single-template fits. 
The statistical independence of these two GeV extended sources allows us to robustly extract spectral and spatial GeV gamma-ray information from the Ridge.   
There have been three main proposals for the origin of the spherically symmetric emission: an unresolved population of millisecond pulsars, dark matter pair annihilation, and CRs interacting with the interstellar medium and radiation field.  Some examples of recent articles discussing the pros and cons of the different proposals are~\citep{MaciasGordon2014,Abazajian2014,Daylan:2014,gomez2013,Lacroix:2014,Bringmann:2014lpa,Cirelli:2014lwa,cholis2014,CarlsonProfumo,
Petrovic:2014uda,Yuan:2014rca,Cholis:2014lta}. 

In this study, we focus on understanding the GeV  and TeV Ridge component, while, at the same time, accounting
for the spherically symmetric generalized NFW component and modeling its spectrum empirically with 
a best-fit log parabola parametric form.

For the Ridge template, we revisit the non-steady-state hadronic model and demonstrate that diffuse GeV-TeV gamma-rays radiation as well as radio continuum emission from the Ridge can be explained by  a flare-like injection of high energy protons  in the surrounding dense gas environment. 
We also update the steady-state models of \cite{Crocker2011b} to include the GeV Ridge data.

\section{Diffuse GeV data from the Ridge region}
\label{sec:diffuseGeVdata}

\begin{figure}
\begin{center}

\begin{tabular}{c}

\includegraphics[width=1.0\linewidth]{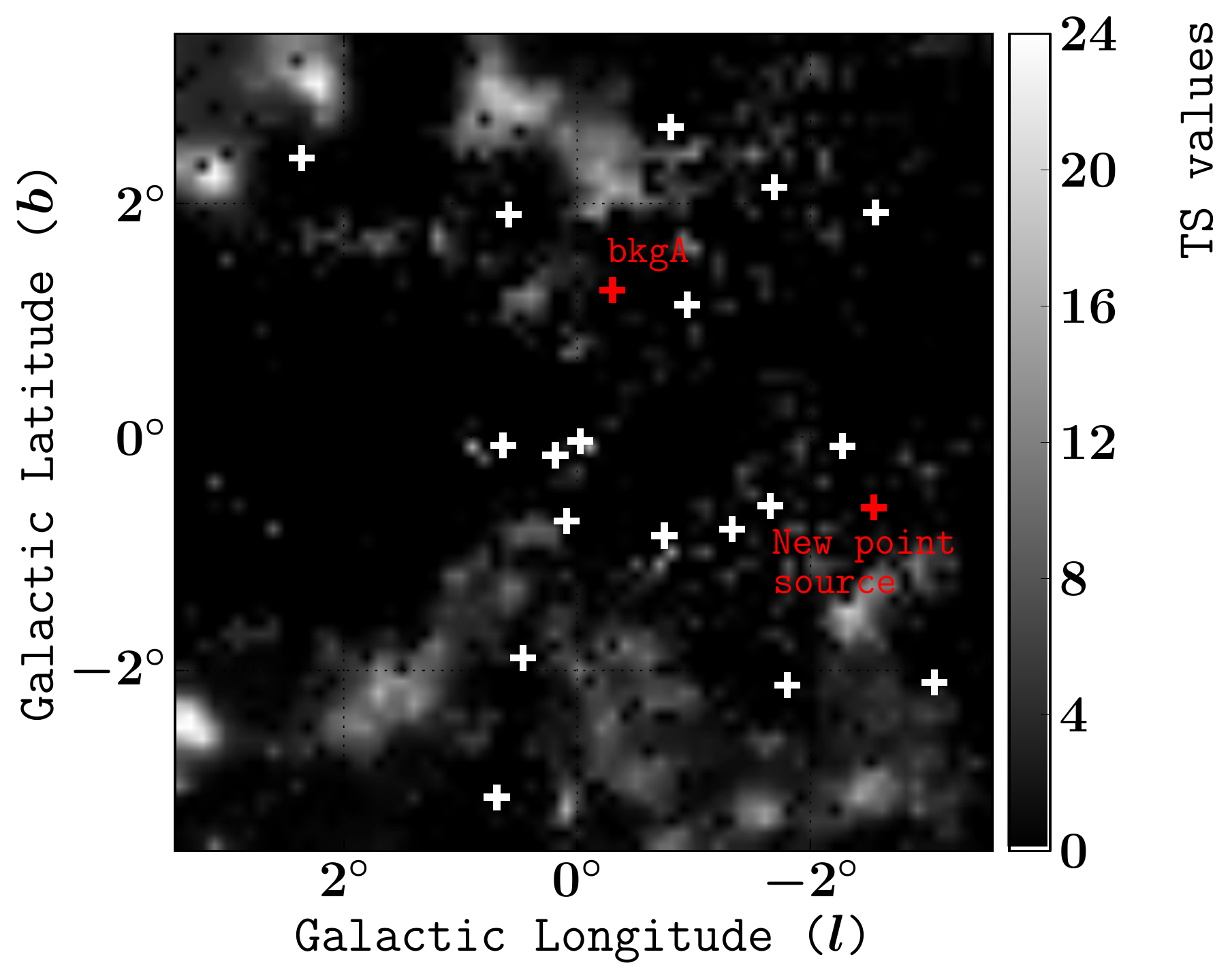}  
\end{tabular}

\caption{ \label{fig:tsmaps}Residual test statistics ($TS$) map in the energy range 200 MeV$-$100 GeV for our best-fitting model of the GC region. See Sec.~\ref{subsec:fermifit} for details on the model considered. This region spans $7^{\circ} \times 7^{\circ}$ centered on Sgr A$^{\star}$ and pixels have dimensions of $0.1^{\circ} \times 0.1^{\circ}$.  White crosses show 2FGL point-sources present in the region~\citep{2FGL}, two red crosses denote two recently detected gamma-rays sources~\citep{Yusef-Zadeh2013, MaciasGordon2014}. As all pixels are below $TS=25$ this indicates the model fits well.
 }   
\end{center}
\end{figure}

 If high-energy gamma rays from HESS Ridge TeV source are produced by the decay of neutral mesons (mostly $\pi^{0}$'s) resulting from hadronic interactions of CRs with interstellar matter, it is plausible that the GeV and TeV emission from the Ridge source originates from the same population of CR particles. This  suggests that extended emission should be detectable at GeV energies. It is expected that analyses of the GeV counterpart of this TeV source will help to single out the emission mechanisms producing high energy photons from the HESS field.

\begin{figure*}
\begin{center}

\begin{tabular}{c}
\begin{minipage}[t]{0.5\linewidth}
\includegraphics[width=1\linewidth]{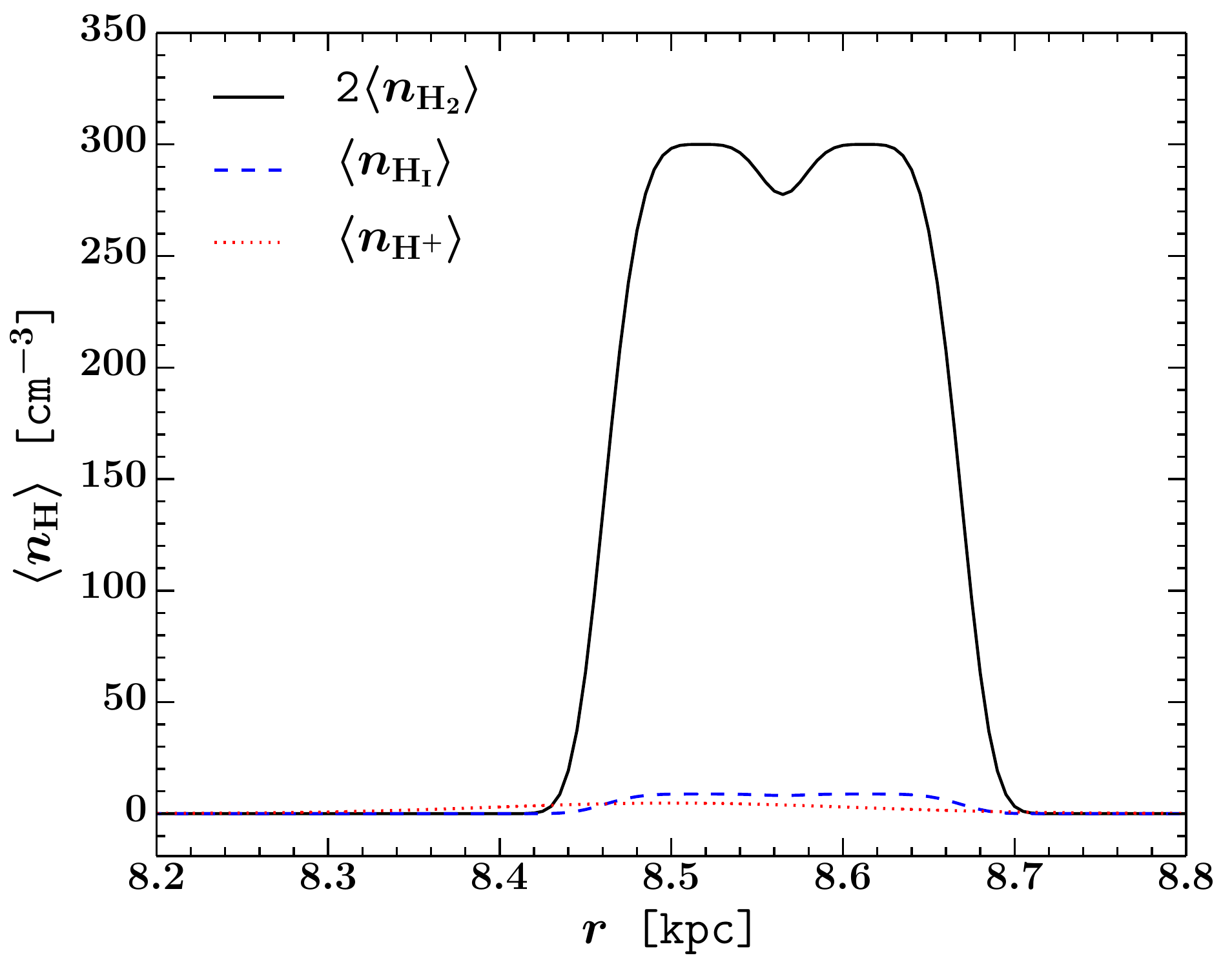}
\end{minipage}
\begin{minipage}[b]{0.5\linewidth}
 \includegraphics[width=1\linewidth]{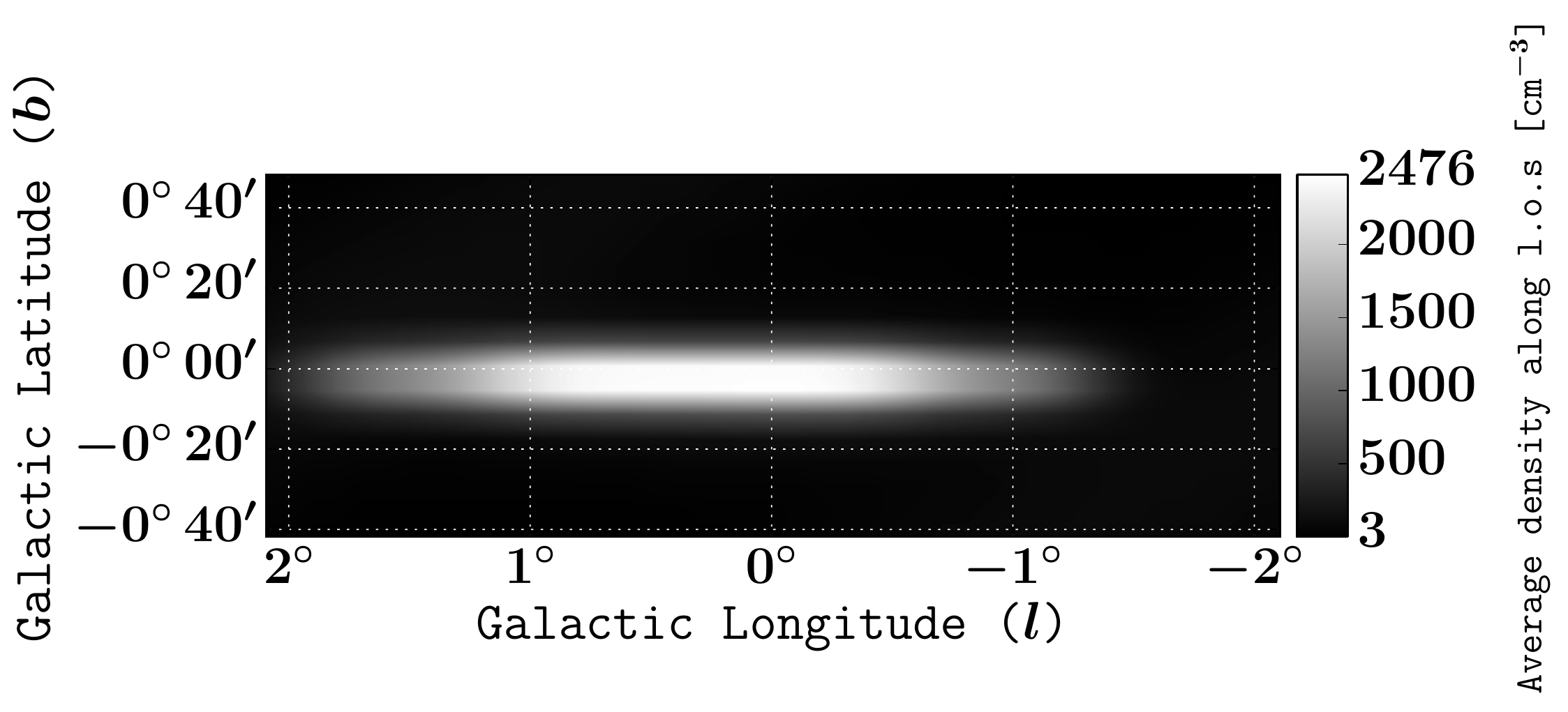} 
\vspace{1.7cm}
\end{minipage}
\end{tabular}

\caption{ \label{fig:Gasmap} Panels shown are obtained from data taken from~\citet{Ferriere}. \textit{Left:} Average densities of interstellar gas as functions of distance $(r)$ along the line of sight passing between us and the GC; Molecular hydrogen is shown with a black solid line, atomic hydrogen with a blue dashed line and ionized hydrogen atoms with a red dotted line. \textit{Right:}  Total space-averaged density of interstellar hydrogen nuclei along the line of sight of the innermost 3 kpc.
}   
\end{center}
\end{figure*}

\subsection{GeV gamma-ray observations of the Ridge}
\label{subsec:fermiobsevations}

The Fermi-LAT telescope detects gamma rays from 20 MeV to more than 300 GeV using particle physics technology~\citep{2009ApJ...697.1071A}. This instrument operates most of the time in continuous sky-survey mode, observing the entire sky every $\sim 3$ hours. We accumulated Pass-7 data taken within a squared region of $7^{\circ}\times7^{\circ}$ centred on Sgr A$^{\star}$ in the first 45 months of observations over the period August 4, 2008$-$June 6, 2012\footnote{Pass-7 data has been superseded by reprocessed Pass-7 (\textsc{Pass7-Rep}). However, 193 weeks of Pass-7 data are still available at \url{http://heasarc.gsfc.nasa.gov/FTP/fermi/data/lat/weekly/p7v6/photon/}. }. 
We
kept only the \texttt{SOURCE} class events, which have a high probability of being photons of astrophysical origin. In order to limit the contamination from the Earth's atmospheric gamma-rays emission, we selected events with measured arrival directions within 100$^{\circ}$ of the local zenith, taken during periods when the LAT rocking angle was less than 52$^{\circ}$. The angular resolution of Fermi-LAT depends on the photon energy, improving as the energy increases~\citep{2009ApJ...697.1071A}.

 We also only selected events between 200 MeV$-$100 GeV without making any distinction between \textit{Front} and \textit{Back} events. Below 200 MeV the angular resolution is poor and source confusion could introduce a large bias, whereas above 100 GeV it is limited by low photon statistics. The sources spectra was computed using a binned likelihood technique~\citep{2FGL} with the \textit{pyLikelihood} analysis tool\footnote{\url{http://fermi.gsfc.nasa.gov/ssc/data/analysis/documentation/}}, and the energy binning was set to 24 logarithmic evenly spaced bins.

\subsection{ Fermi-LAT analysis methods}
\label{subsec:fermifit}

The spatial and spectral features of a source are intrinsically correlated. An inaccurate spatial model would affect the source spectra and vice versa. In~\citet{MaciasGordon2014} some of us showed evidence for an extended gamma-ray emitting source that is the GeV counterpart of diffuse emission detected by HESS and some radio telescopes~\citep{Aharonian:2006au,Crocker2011b,Yusef-Zadeh2013}. The fit included all 2FGL~\citep{2FGL} point-sources present in the region of interest as well as standard diffuse Galactic and isotropic extra-galactic models \texttt{gal$_{-}$2yearp7v6$_{-}$v0.fits}  and  \texttt{iso$_{-}$p7v6source.txt} respectively.

 \cite{MaciasGordon2014} performed a set of maximum likelihood fits using templates for the Ridge source. Either a 20-cm  continuum emission map \citep{Yusef-Zadeh2013} or a HESS residuals \citep{Aharonian:2006au}  map were used.
{\color{Black} The 20-cm template was based on  Green Bank Telescope continuum emission data  which measures nonthermal and thermal plasma distributions \citep{law2008,Yusef-Zadeh2013}. 
	The Ridge template has a very different morphology to the diffuse Galactic background (DGB) template which is modelled as a linear combination of gas column-densities,
an inverse Compton intensity map, and an isotropic intensity
that accounts for the extragalactic background and the residual instrumental
background \citep{DGB}. The DGB column densities are based on radio and millimetre
line surveys of HI and CO. Within 10$^\circ$ of the GC, these line surveys lose kinematic resolution.   When constructing the DGB template, the distributions
of the gas along the line of sight were interpolated from their distributions
at longitudes just outside 10$^\circ$ of the GC. Also the spectrum of the cosmic rays is assumed to be uniform on kpc scales in the 
DGB  template. The combination of these factors results in the DGB having a morphology that is not degenerate with the Ridge template obtained from the 20cm continuum emission template or the HESS residuals.}

 Although it was found that the best fit model results from a combination of the 20-cm map with point sources ``the Arc'' (2FGL J1746.6-2851c) and ``Sgr B'' (2FGL J1747.3-2825c), it is possible that these two point sources also result from the interaction of CRs with molecular gas and so as to 
maximize the  
Ridge signal, they were not included~\citep{Yusef-Zadeh2013,MaciasGordon2014}.     

\cite{MaciasGordon2014} confirmed evidence for a spherically symmetric extended source, whose spectrum and morphology is consistent both with emission from millisecond pulsars or dark matter annihilation. Most prominent is the fact that the spatial extension and spectrum of the Ridge source was shown to be robustly independent of the spherical gamma-ray source. Although, additional complications 
 due to systematic uncertainties  {\color{black} arise}  in the DGB model~\citep{2FGL}, the analysis in~\citet{MaciasGordon2014} assessed such systematic uncertainties concluding that these are in the vicinity of  20\%.  For this work we use similar analysis methods and assume the same estimates for the systematic uncertainties in the DGB.

Since the 2FGL catalog~\citep{2FGL} was constructed with 2 years of data, while our dataset comprises almost 4 years, we searched for additional point sources within the ROI by constructing maps of residual significance after subtracting all known sources in the region. This was done with the \textsf{gttsmap} tool, a specialized routine provided with the \textit{Fermi} analysis software that computes test statistic ($TS$) images. The $TS$ gives a measure of the significance of adding a source to a model, defined as $TS = 2 \log(\mathcal{L}_1 /\mathcal{L}_0)$ where $\mathcal{L}$ is the  likelihood, and the subscripts 0 and 1 refer to the original model and a model with an additional source, respectively. We illustrate in Fig.~\ref{fig:tsmaps} the adequacy of the astrophysical model used here once two new faint point-sources plus the Ridge and spherical extended source are included. 

We performed
a curvature analysis~\citep{2FGL} intended to determine whether the Ridge spectrum
can be fitted with a power law or needs a curved spectrum such as a broken power law.
 We found that the broken power-law spectrum is preferred at about $7\sigma$ significance.

We also extracted the gamma-ray spatial distribution from the Ridge by performing a longitudinal analysis that mirrors the one done by the HESS collaboration for the same site~\citep{Aharonian:2006au}. We obtained a residual counts map of the Ridge in the energy range $0.2-100.0$ GeV by subtracting out all of our best-fitting sources from the raw counts map, except for the 20 cm map source. The background level was estimated using events from the regions $0.8^{\circ}< \abs{ b } <1.5^{\circ}$ and is somewhat longitude dependent. Counts obtained at each different bin, $0.2^{\circ}$ wide, were thus background subtracted.

\section{Gas Density Maps of the inner 200 parcecs region of the Galactic Center}
\label{sec:gas}

\begin{figure}
\begin{center}

\begin{tabular}{c}

\includegraphics[width=1.0\linewidth]{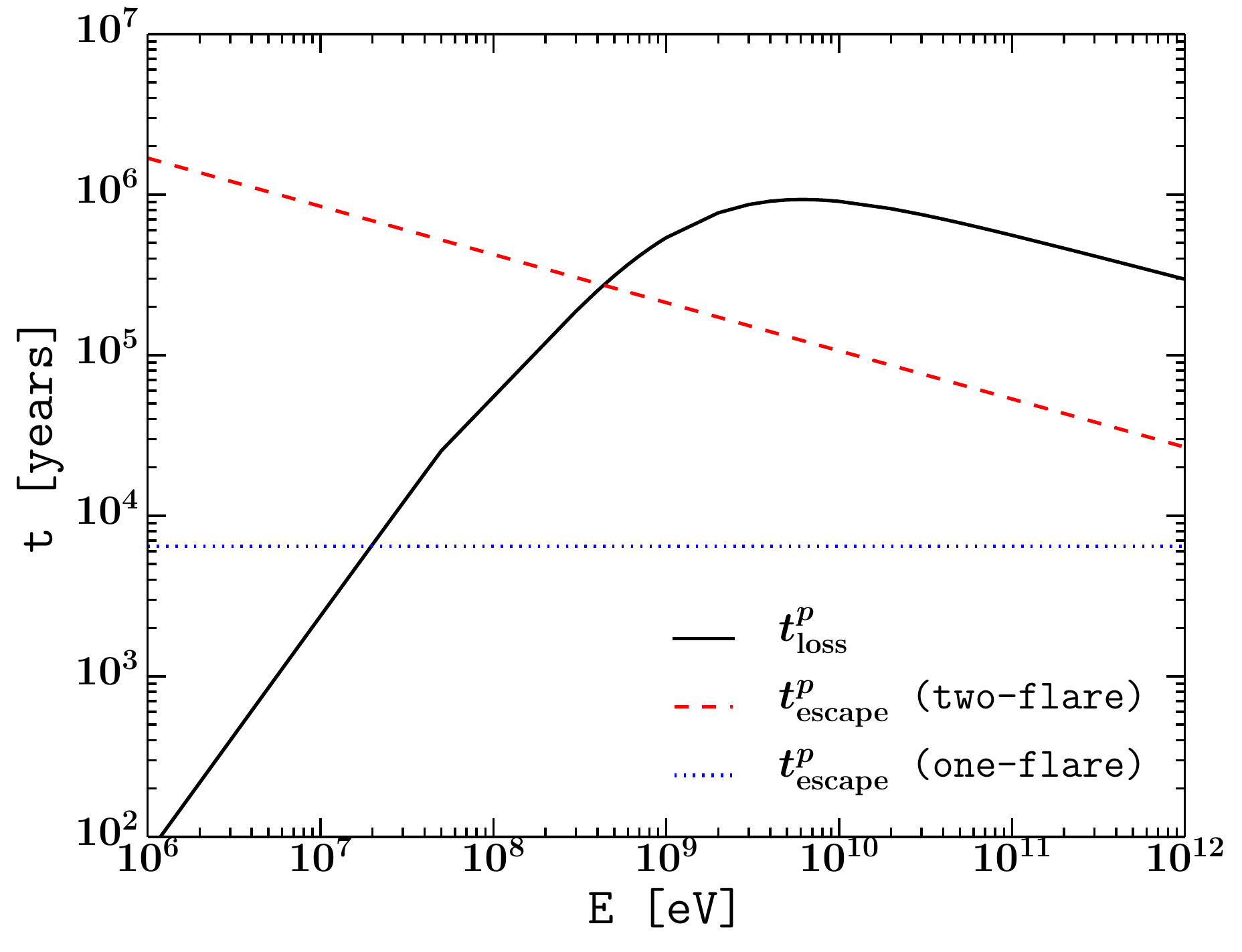}  
\end{tabular}

\caption{ Proton energy-loss time-scales ($t^p_{\rm loss}=\E/\frac{d\E}{dt}$) in the Ridge region. The total average hydrogen density used for this calculation is $\overline{\langle n_{H}\rangle} = 109.3$ cm$^{-3}$~\citep{Ferriere}. 
{\color{Black}	From Eq.~(\ref{Rdiff}), the escape time is $t_{\rm escape}=(\Delta r)^2/(2 D)$ where for $\Delta r$ we use the Ridge dimensions 0.8$^\circ$ which corresponds to about 118.7 pc for the GC being 8.5 kpc away. We include both the two flare of Sec.~6.1 ($D=10^{28}(E/{\rm GeV})^{0.3}$cm$^2$s$^{-1}$) and single flare of Sec.~7.1 ($D=0.33*10^{30}$cm$^2$s$^{-1}$) diffusion coefficients.}
 \label{fig:tploss}
	}   
\end{center}
\end{figure}
In order to perform realistic simulations of hadronic gamma-ray emission from the GC a detailed knowledge of the spatial distribution of interstellar gas in the region is necessary. Here, we employ the model of ~\citet{Ferriere}, valid for the innermost 3.0~kpc of our Galaxy. This model provides three-dimensional (3D) hydrogen space-averaged densities maps that best fits the observational data while being entirely consistent with theoretical predictions. This will prove pivotal to our present study as it enables us to create fully 3D CR propagation simulations.   

Fig.~\ref{fig:Gasmap}\textit{-left} illustrates the radial variation of the column densities of molecular, atomic and ionized gases. The most abundant material in our region of interest is molecular hydrogen $\langle n_{H_{2}} \rangle$. This component forms a Galactic structure known as the CMZ$-$an asymmetric layer of predominantly molecular gas that encompasses the region defined by $-1.5^{\circ}\lesssim l\lesssim 2.0^\circ$ and $\abs{b}\leq0.3^\circ$ around Sgr A$^{\star}$. The right hand side panel of Fig.~\ref{fig:Gasmap} is obtained by computing the mean value of the total gas density
\begin{equation}
\label{eq:n_H}
\langle n_{H}\rangle = 2\langle n_{H_{2}} \rangle + \langle n_{H_{I}} \rangle + \langle n_{H^{+}} \rangle,
\end{equation}
 along the line of sight direction. From this we can thus expect that one or more CR accelerators injecting protons into the medium will generate (given an appropriate choice of diffusion parameters) gamma-ray distributions that approximately follow the density of interstellar gas displayed in Fig.~\ref{fig:Gasmap}.

\section{Multiwavelength Modeling}
\label{sec:modeling}

The origin of extended gamma-ray emission from the direction of the Ridge is not yet firmly established. Despite the fact that the region of emission ($\abs{ l }< 0.8^{\circ}$, $\abs{ b }<0.3^{\circ}$) and spectra have been detected with great accuracy by HESS, these observations can be well explained by more than one  mechanism~\citep{Crocker2011b,Yusef-Zadeh2013,YoastHullGallagherZweibel2014}. Interestingly, the HESS team interpreted the breakdown in the correlation between the diffuse TeV emission and the molecular hydrogen density as an indication of a non-steady-state phenomena. Such a model however must be carefully evaluated for consistency in light of recent measurements~\citep{Yusef-Zadeh2013,MaciasGordon2014} at lower energies. Here we revisit the non-steady-state gamma-ray production scenario related to past activity of Sgr A$^{\star}$ or possibly supernova remnants in its immediate vicinity.

\subsection{Computation of gamma ray and $e^{\pm}$ spectra in a non-steady-state Hadronic emission model}

~\cite{Liuetal} argued that protons interacting resonantly  with turbulent electromagnetic fields in the accretion torus of Sgr A$^{\star}$ can undergo stochastic acceleration. A significant fraction of these protons may diffuse out of the system and enter the interstellar medium~\citep{Neronov, Liuetal}. Of high importance to our approach is the time evolution of the proton injection. This will be dominated by past flares from Sgr A$^{\star}$ (or equivalently a few supernova explosions). The propagation of runaway protons boosted by such  impulsive activity is modeled with a phenomenological solution to the diffusion equation~\citep{Syrovatskii} that prevents CRs from reaching superluminal velocities~\citep{Aloisio}.

We describe the spatial and energy distribution of protons at a given time~\citep{Aloisio} by
\begin{eqnarray}
\label{eq:solutiontoDiff}  \nonumber
\frac{dn_p(\E,r)}{d\E} &=& \frac 1{4 \pi c r^2} \int \limits_{\xi_{{\rm min}}}^{\xi_{{\rm max}}} d\xi\;
\frac{Q[\Eg(\E,\xi)]}{(1-\xi^2)^2} \; \xi\;
\frac{\alpha(\E,\xi)}{K_1[\alpha(\E,\xi)]}\; \\
& \times & \exp\left[-
\frac{\alpha(\E,\xi)}{\sqrt{1-\xi^2}} \right] \qquad \left[ {\rm cm^{-3}\;eV^{-1}}\right],
\end{eqnarray}
where $K_1(x)$ is the modified Bessel function, $c$ the speed of light, $\Eg$ is the generation (or initial) kinetic energy of a proton, $\E$ is the cooled proton kinetic energy at time $t$ and $Q[\Eg(\E,\xi)]$ is the injection spectrum.

For convenience, the dimensionless variable $\xi = r/ct$ is substituted for the time $t$ . The integration limits corresponding to our flare scenario are   
\begin{eqnarray}
\xi_{\rm min} = \frac{r}{ct_0}  \quad {\rm and} \quad \xi_{\rm max} ={\rm Min}\left[\frac{r}{c(t_0-\Delta t)},1\right],
\end{eqnarray}
where $t_0$ represents the time that has lasted since the source switched on and $\Delta t$ corresponds to the duration time of the flare event. 

The dimensionless function $\alpha(\E,\xi)$ in Eq.~(\ref{eq:solutiontoDiff}) is defined as follows
\begin{eqnarray}
\label{eq:substitute}\nonumber
\alpha(\E,t) =  \frac{c^2t^2}{2\lambda[\Eg(\E,t)]}\mbox{ with } \\ \lambda(\E,t) = \int_{\E}^{\Eg(t)} d \E'\;
\frac{D(\E')}{b(\E')},
\end{eqnarray}
where $D(\E)$ is the diffusion coefficient. Also, the total energy loss rate of CR protons, given by
\begin{equation}
\label{eq:lossfunction}
b(\E)=-\frac{d\E}{dt},
\end{equation}
is mainly due to pion production, Coulomb losses and ionization interactions. In this section, we consider full energy loss expressions provided in Eq.(5.3.58) of \citet{Schlickeiser}. Making use of these explicit functions for $b(\E)$, the generation energy $\Eg(\E,t)$ can be readily computed by integrating Eq.~(\ref{eq:lossfunction}). We display in Fig.~\ref{fig:tploss} the relevant time-scales in which proton energy losses are significant.

\begin{figure}
\begin{center}

\begin{tabular}{c}

\includegraphics[width=1.0\linewidth]{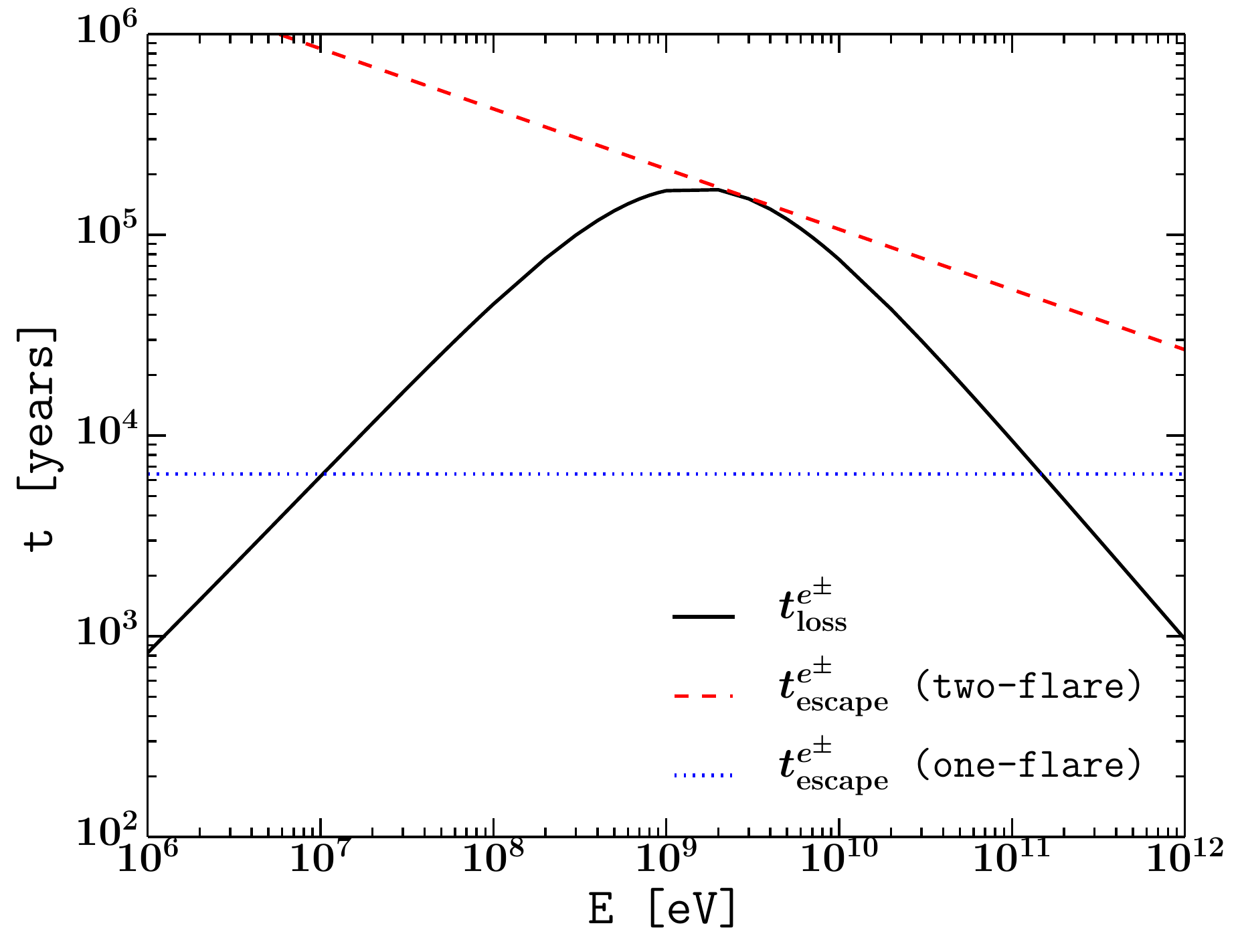}  
\end{tabular}

\caption{ \label{fig:te+-loss} Total loss time scale  ($t^{e\pm}_{\rm loss}=E/\frac{dE}{dt}$) for electrons and positrons in the Ridge region ($\overline{\langle n_H\rangle}= 109.3$ cm$^{-3}$ and $B = 119\;\mu$G). Energy losses considered in this work are: ionization, bremmstrahlung, synchrotron and inverse Compton. See the caption of Fig.~\ref{fig:tploss} for the $t_{\rm escape}$ definitions.}   
\end{center}
\end{figure}
  
We assume here, for simplicity, that $D(\E)$ is independent of position, and given by 
\begin{eqnarray}
\label{eq:diffusionConstant}
D(\E) = 10^{28}\;\left(\frac{\E}{\rm 1\; GeV  } \right)^{\beta}\times \kappa \qquad [{\rm cm^2\; s^{-1}}],
\end{eqnarray}
where $\beta$ and $\kappa$ are 
free parameters. 
 For the initial spectrum of protons $Q[\Eg(E,t)]$  we use a power-law with an exponential cutoff
\begin{eqnarray}
\label{eq:injectionSpectrum}\nonumber
Q[\Eg(E,t)] &=&K\;  \left(\frac{\Eg}{ 1\;{\rm GeV} }\right)^{-\Gamma}\exp\left(-\frac{\Eg}{ 100\;{\rm TeV} }\right)\; \\&&   \left[ {\rm eV^{-1}\; s^{-1}}\right],
\end{eqnarray}
where 
$K$ is a normalization constant, $\Gamma$ is the spectral slope.
For our calculations, we actually used an exponential cut-off power law in momentum as that is the generic prediction from shock acceleration. But in practice we found
this only makes a negligibly small change, for injection energies larger than the proton mass,
 compared to having an exponential cut-off in kinetic energy as employed by \cite{Chernyakova:2011zz}. 
 Both $K$ and $\Gamma$  are left free in our parameter estimations. 
 We note that  in Eq.~(2) of~\citet{Chernyakova:2011zz} there is a typo
in that the injection spectrum should depend on the generation rather than the observed energy. %

Finally, the differential emissivity of secondary particles (photons, electrons, or positrons) resulting from collisions between a distribution of injected protons $d n_p(\E,r)/d \E$ (as obtained in Eq.~(\ref{eq:solutiontoDiff})) into ambient hydrogen of Ridge average density
 $\overline{\left\langle n_H \right\rangle}= 109.3$ cm$^{-3}$ is calculated as 
\begin{eqnarray}
\label{eq:sourcefunction}\nonumber
q_{\gamma, e^{\pm}}(E,r) &=& c \;\overline{\left\langle n_H \right\rangle} \int^{\infty }_{0}d\E'\, 
\frac{d n_p(\E',r)}{ d\E'} \;\frac{d\sigma_{\gamma, e^{\pm}}(E,\E')}{dE}\\
&& \left[\rm secondaries\; eV^{-1}\; cm^{-3}\;s^{-1} \right],
\end{eqnarray}
where $d\sigma_{\gamma, e^{\pm}}(E,\E')/ dE$ is the differential cross-section which we are taking to be non-zero only for kinematically allowed energies. This is provided by~\citet{Kamaeetal} in the form of interpolation functions for several different final state particles. The actual spectrum of secondaries is then obtained by integrating the source function along the line of sight and over the angular area ($\Delta \Omega$) of the Ridge 
\begin{eqnarray}\nonumber
\label{eq:spectra}
\frac{dN_{\gamma, e^{\pm}}(E)}{dE} &=& \int_{\Delta \Omega}d\Omega \int^{+\infty }_{0}ds\;q_{\gamma, e^{\pm}}(E,r)\\ && \left[\rm secondaries\; eV^{-1}\; cm^{-2}\;s^{-1} \right],
\end{eqnarray}
where we use the conventional coordinates transformation
\begin{equation}
\label{eq:radius}
r=\sqrt{R^2_{\odot}-2sR_{\odot}\cos(b)\cos(l)+s^2}\;,
\end{equation}
with $s$ varying along the line-of-sight path, $b$ and $l$ the Galactic coordinates and $R_{\odot}=8.5$ kpc is the distance from the solar system to the GC.

As we have seen, the gamma-ray emission is proportional to the pion production rate, and because the lifetime of pions is extremely short, the location of the gamma-ray emission is essentially that of the proton scattering. 
To construct the spatial morphology of our gamma-ray predictions, we first multiply the three dimensional proton distribution (Eq.~(\ref{eq:solutiontoDiff})) with the three dimensional spatially varying gas map $ \langle n_H\rangle$ we obtained from \cite{Ferriere}. We then take the resulting three dimensional distribution and perform  line of sight integrations from the solar system position to construct the two dimensional map of the predicted spatial morphology for the gamma-ray maps
that were tested against GeV and TeV data. Our methods for performing the GeV spatial fits are explained in~\citet{gordonmacias2013}.         

\subsection{Synchrotron emission from $e^{\pm}$ of hadronic origin}

\begin{figure*}
\begin{center}

\begin{tabular}{cc}

\includegraphics[width=0.5\linewidth]{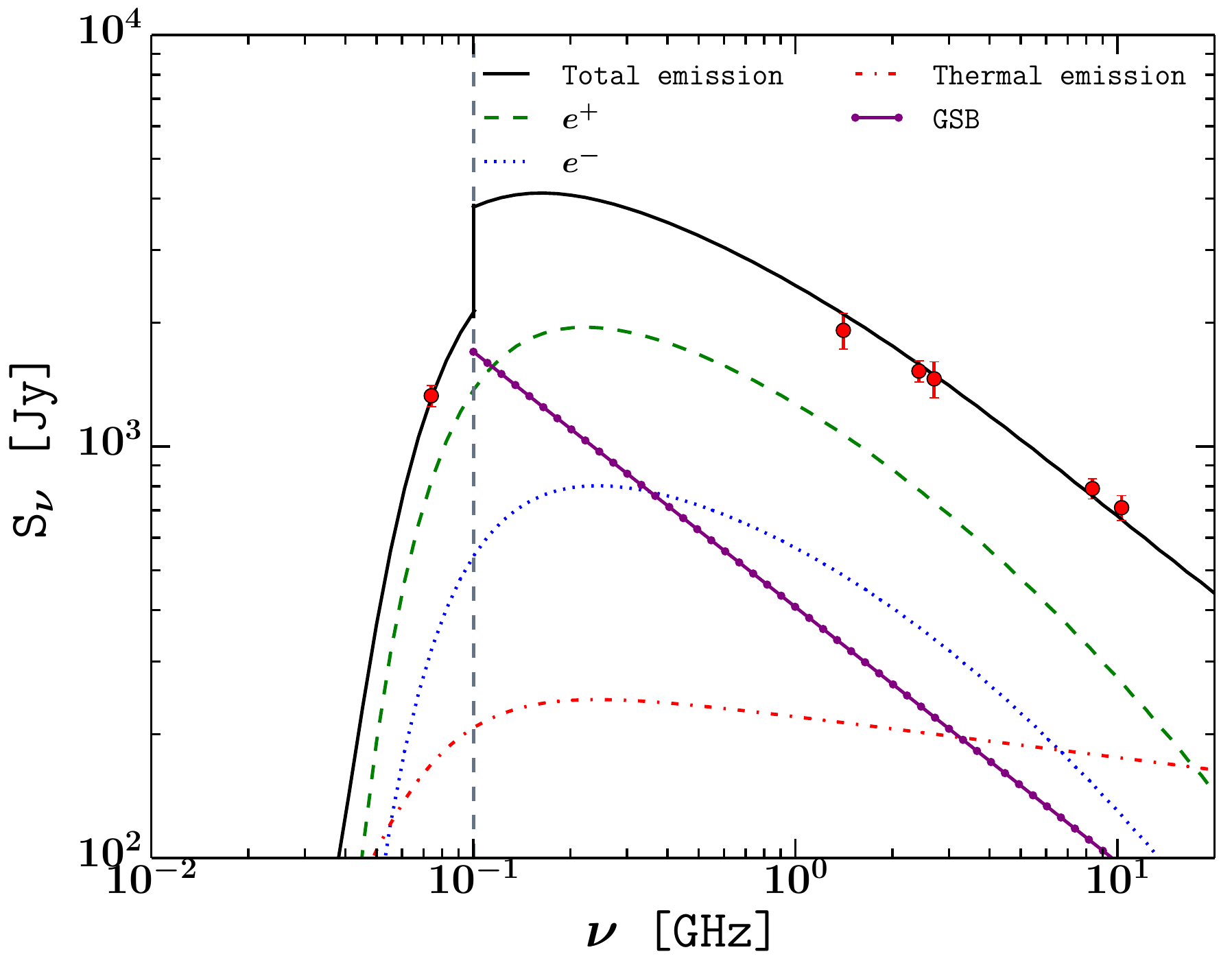} & \includegraphics[width=0.5\linewidth]{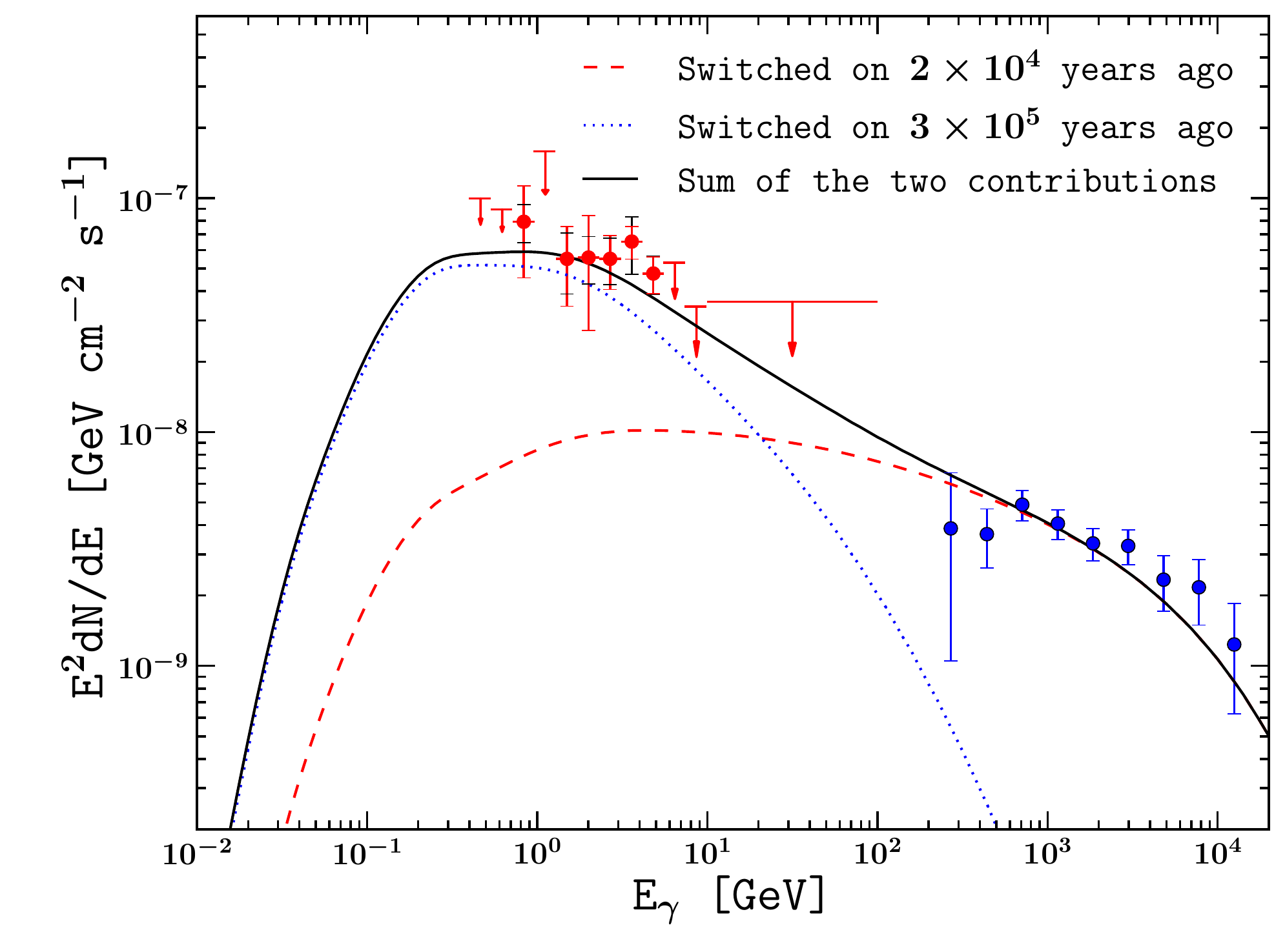}\\
\end{tabular}

\caption{ \label{fig:spectralfitBlackHole}  HESS Ridge region  spectrum for the best-fitting, non-steady-state, Kolmogorov-type turbulence,  hadronic scenario. The model consists of two independent flares that might have occurred $2\times 10^4$ and $3\times 10^{5}$ years ago and lasted for approximately $10$ years. Fitted parameters are summarised in Table~\ref{tab:FitGeVTeVFlareModel}. We assume the gas maps provided in~\citet{Ferriere}. Data are from HESS diffuse~\citep{Aharonian:2006au}, Ridge Fermi-LAT~\citep{MaciasGordon2014} and radio observations~\citep{Crocker2011b}. 
We display modeled synchrotron emission from CR electrons and positrons of hadronic origin at radio wavelengths. Total radio emission is the combination of thermal and non-thermal emission, both of which are affected by free-free absorption. We also include a Galactic synchrotron background (GSB) component for $\nu>0.1$ GHz to account for foreground and background~\citep{Crocker2011b}. Notice, that the unphysical step at 100 MHz is explained by the fact that the first datum is interferometric and does not receive a contribution from the large angular scale, line-of-sight GSB emission. 
 }   
\end{center}
\end{figure*}

\begin{figure*}
\begin{center}
\begin{tabular}{cc}
\includegraphics[width=0.5\linewidth]{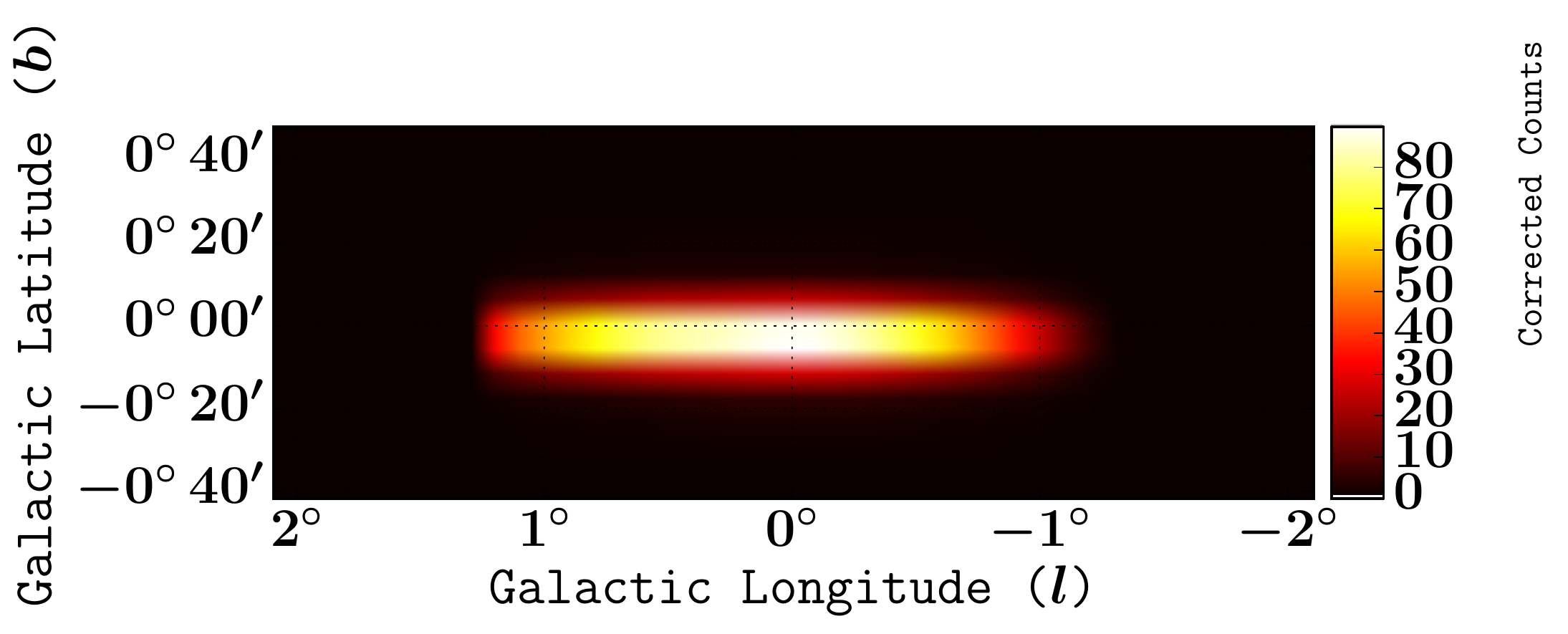} & \includegraphics[width=0.5\linewidth]{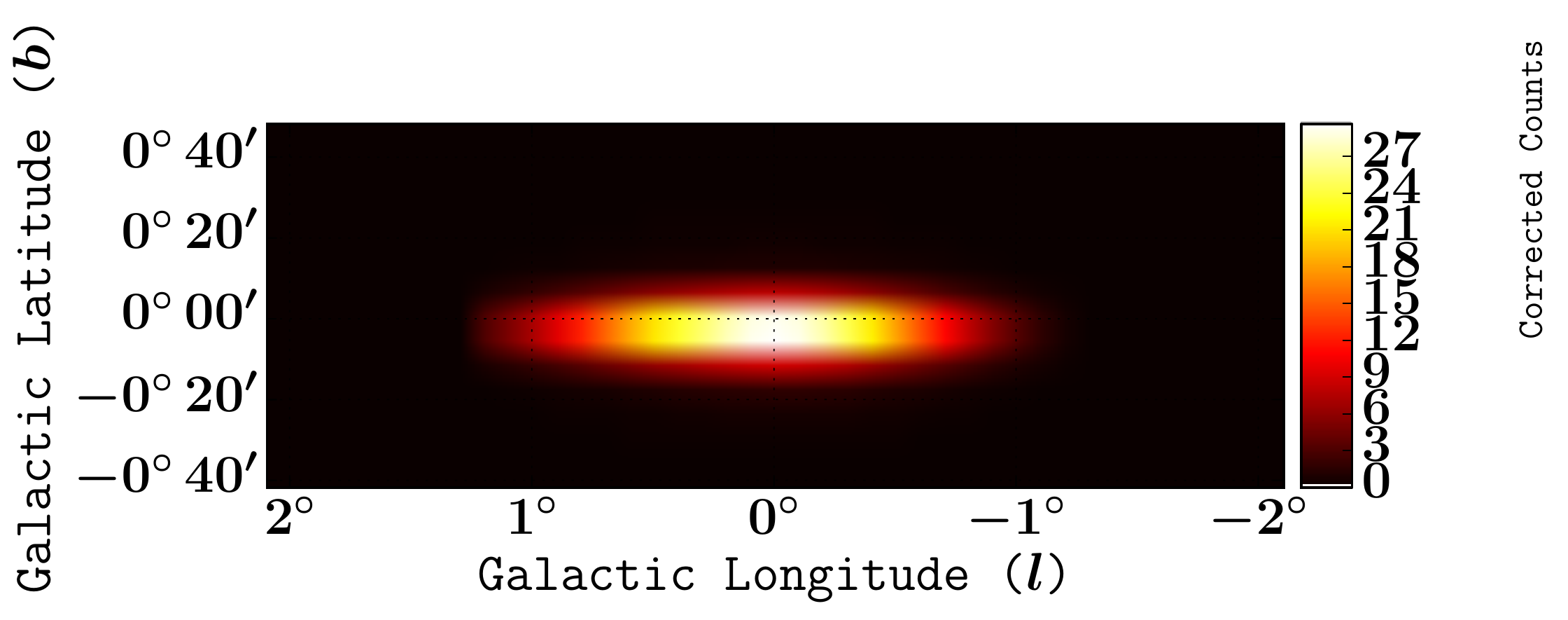}\\
\includegraphics[width=0.5\linewidth]{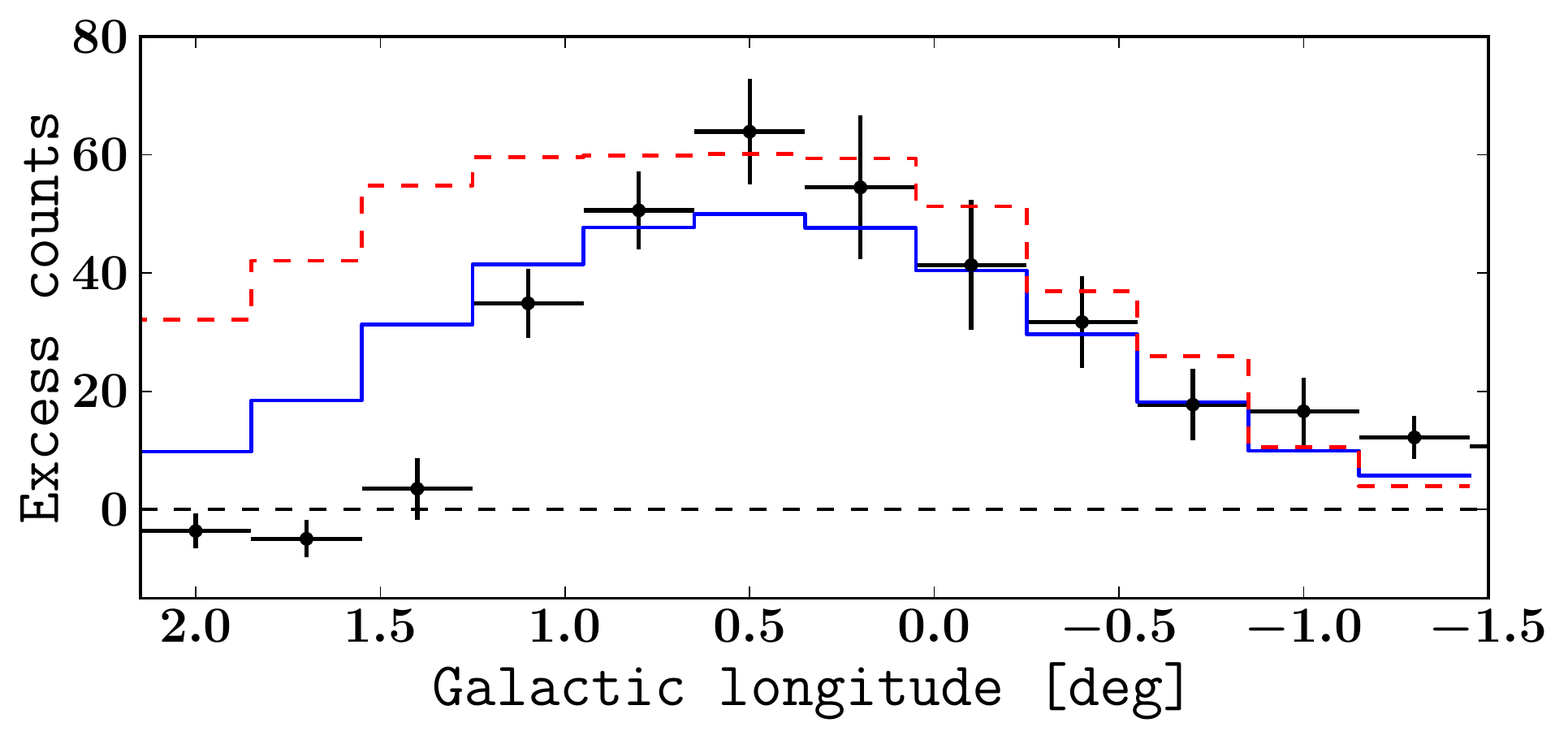} & \includegraphics[width=0.5\linewidth]{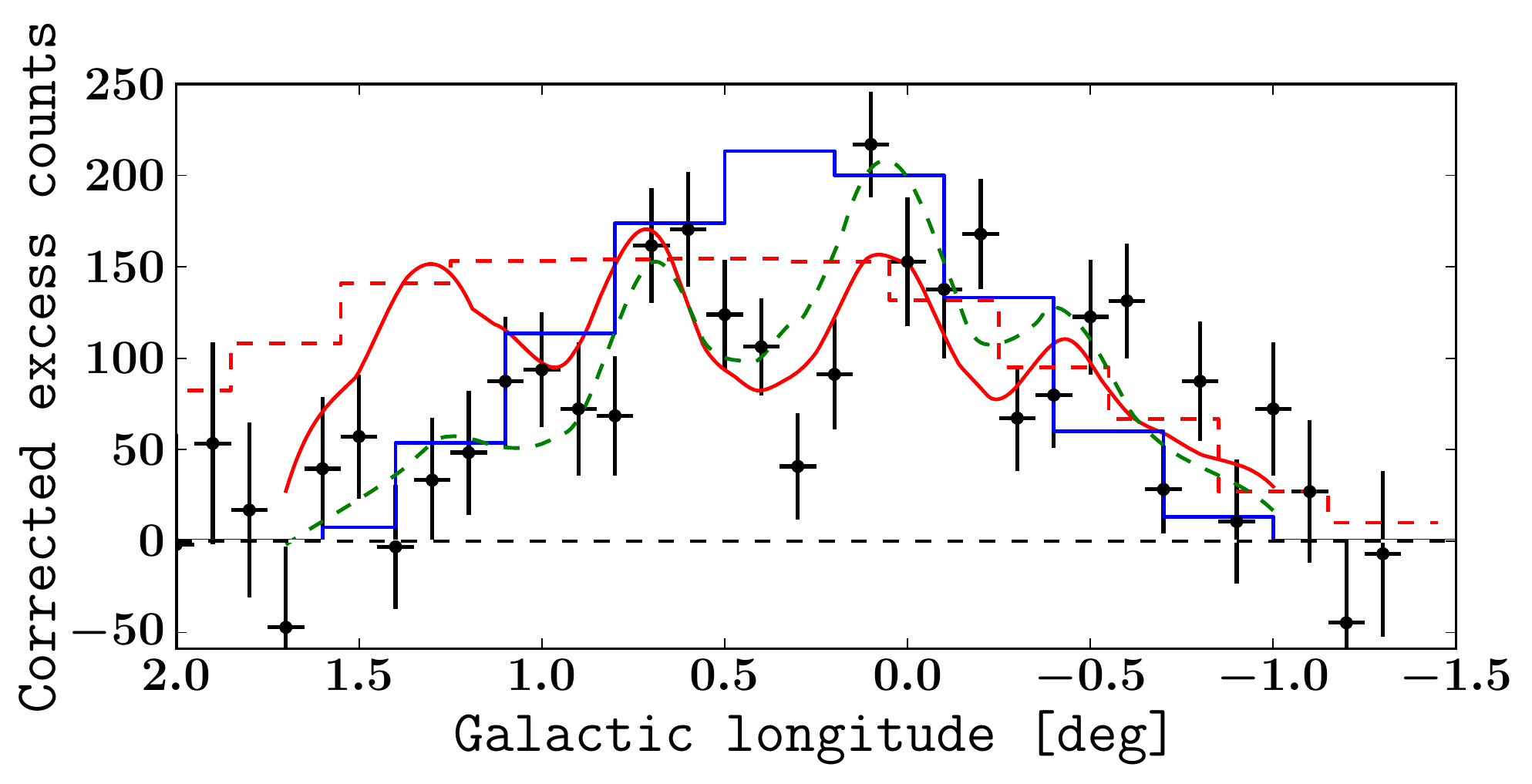}
\end{tabular}

\caption{ \label{fig:spatialfitBlackHole} Brightness profile of the Ridge in $0.2-100.0$ GeV (left column) and $0.27-12.5$ TeV (right column) energy ranges for our best-fitting non-steady-state hadronic model. See Sec.~\ref{sec:results} for a description of the models. \textit{Top panel:} Two-dimensional distributions of gamma-ray emission from our best fit GeV flare (left) and TeV flare (right) models. Below  each 2D counts map we display its corresponding longitudinal profile. \textit{Bottom panel:} Gamma-ray counts versus longitude. The background level was estimated using events from the regions $0.8^{\circ}< \abs{ b } <1.5^{\circ}$. Counts obtained at each different bin, $0.2^{\circ}$ wide, are background subtracted. 
The red dashed histograms show the density of molecular gas as obtained from~\citet{Ferriere}, the red solid curve shows target gas as traced by CS emission~\citep{Aharonian:2006au}. The blue histograms show the longitudinal distribution of photons as predicted by our model at the two different energy ranges. For comparison we show the best-fit obtained in~\citet{Aharonian:2006au} with a green-dashed line.   }   
\end{center}
\end{figure*}

For the Ridge environment, we have a total average hydrogen density of $\overline{\langle n_{H}\rangle} = 109.3$ cm$^{-3}$~\citep{Ferriere} and, as we show below, our fit to the radio data suggests an average magnetic field $B = 200\;\mu$G.

In our non-steady-state hadronic emission model, synchrotron radiation at a frequency of $\sim 1$ GHz will be generated by secondary leptons of energy $E_{e^{\pm}} \simeq {\rm 1\; GeV}$ (see appendix of~\citet{MaciasGordon2014}). As discussed later,
 we consider our 
diffusion coefficient in the  Ridge environment to be $D(E) = 10^{28} \, (E/1 \, {\rm GeV})^{0.3}$ cm$^{-2}$ s$^{-1}$.  
{\color{black} From Eq.~(\ref{eq:substitute}) it can be seen that the diffusive transport scale  is given by $R_{\rm diff} = \sqrt{2\lambda(E,t_0) }$ so that 
\citep{aharonian2004}
\begin{equation}
\begin{array}{lcl}
R_{\rm diff} &=& \sqrt{2\lambda(E,t_0) }\\
 &\approx& \sqrt{2Dt_0}, \mbox{ for } t_0\ll t_{\rm loss}.
\end{array}
\label{Rdiff}
\end{equation}
and $\lambda\approx Dt_0$ when $t_0\ll t_{\rm loss}$. 
{\color{Black} As can be seen from Fig.~\ref{fig:te+-loss}, in the two flare case, the escape time is greater than the electron/positron  energy loss time for all energies.} 
Based on this comparison, we neglect the  diffusive transport in our estimation and can thus solve the diffusion equation in the thick target limit~\citep[e.g.,][]{Crocker:2007da}:
\begin{equation}
\frac{d n_{e^{\pm}}(E,r)}{d E} = 
{\int_{E}^\infty q_{e^{\pm}}(E', r) dE' \over  b_e(E)} \mbox{ ~~ [cm$^{-3}$ eV$^{-1}$]},
\label{eq:diffusionelectrons}
\end{equation}
where $q_{e^{\pm}}(E',r)$ is given by Eq.~(\ref{eq:sourcefunction}) and $b_e(E)=-dE_e(E)/dt$ is the total energy loss rate of electrons, which is calculated using the formulas presented in~\citet{Delahaye} and plotted in 
Fig.~\ref{fig:te+-loss}.
We take into account energy losses due to ionization, bremmstrahlung, synchrotron and inverse Compton (IC). We also assume that electrons and positrons suffer identical energy losses and neglect electron positron annihilation.

\begin{table*}
	
\begin{tabular}{|r|c|c|c|c|c|c|c|c|}
	\hline
\hline 	
Parameter	 & $\Gamma_{\rm GeV}$ & $\Gamma_{\rm TeV}$ &$E_{\rm total,GeV}$[$10^{50}$ erg]  & $E_{\rm total,TeV}$[$10^{50}$ erg] &$B$[$\mu$G]  & Free-free flux density at 10 GHz [Jy] & $\Gamma_{\rm GSB}$&$\chi^2_{\rm min}$/dof \\ 
\hline 
Mean$\pm$error	 & $1.9\pm0.1$ &$1.8\pm0.3$ &$120^{+50}_{-60}$  & $1.5\pm1.0$ &$400\pm300$& $150\pm 50$ & $0.6\pm0.1$& \\ 
	\hline 
	Best fit	 & 2.0 & 1.8 &150  & 1.5 &200 & 180 & 0.6& 16.0/ 21\\
	\hline \hline
\end{tabular} 
\caption{
	\label{tab:FitGeVTeVFlareModel}
	Propagation parameters obtained from the full broad-band spectral
	observations. The column corresponding to $\chi^2$ is obtained from the fit to spectra shown in  Fig.~\ref{fig:spectralfitBlackHole}. The GSB provides an effective extra radio data-point. The mean values are obtained from marginal distributions assuming a non-negative prior for all parameters. The errors correspond to the 68\% confidence intervals. }
\end{table*}

The power per frequency of emitted synchrotron radiation for a single electron with energy $E=\gamma m_ec^2$
is \citep{Schlickeiser};
\begin{equation}
\label{eq:Powerelctron}
 P_{\nu}(\nu,E) = \frac{\sqrt{3}\;e^3\;B\;\sin\alpha}{m_e\;c^2}\;F(x)\qquad \left[\rm erg\; s^{-1}\; Hz^{-1}\right],
\end{equation}
where $e$ is the electron charge in statcoulombs, $\alpha$ is the pitch angle and $x=\nu/\nu_c$. The quantities $\nu_c$ and $F(x)$ are defined as
\begin{eqnarray}
\label{eq:nuc}\nonumber
\nu_c = \frac{3\;\gamma^2\;B\;\sin\alpha}{2\;m_e\;c} \quad {\rm and}\quad F(x) = x\int_x^{\infty}K_{5/3}(\xi)\mathrm{d}\xi\;,\\
\end{eqnarray}
where $K_{5/3}(\xi)$ is the modified Bessel function of order $5/3$ and we assume  $\langle B_{\perp} \rangle =0.78 B$~\citep{Crocker:2007da}. Finally, we calculate the non-thermal synchrotron radio emission $S_{\rm NT}$, for a given distribution of secondary leptons by using the following expression:
\begin{eqnarray}\nonumber
S_{\rm NT}&=&\int_{0}^\infty \int_{m_e c^2}^{\infty}ds\;dE\;P_{\nu}(\nu,E)\;\frac{dn_{e^{\pm}}(E,r)}{dE}\\ &&\left[{\rm Jy}\;{\rm sr^{-1}} \right].
\label{eq:synchrotron}
\end{eqnarray}

\begin{figure*}
\begin{center}
\begin{tabular}{cc}
\includegraphics[width=0.5\linewidth]{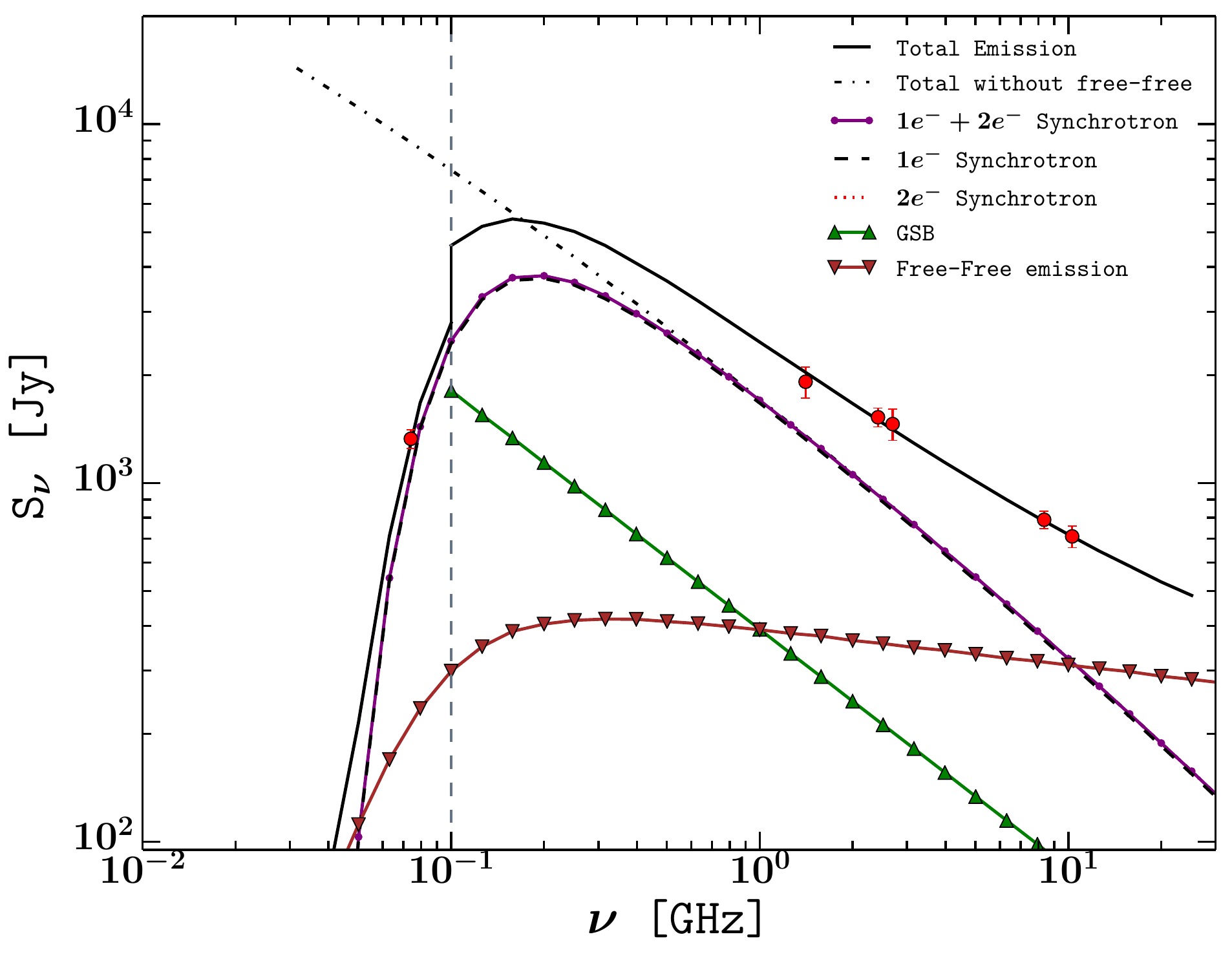} & \includegraphics[width=0.5\linewidth]{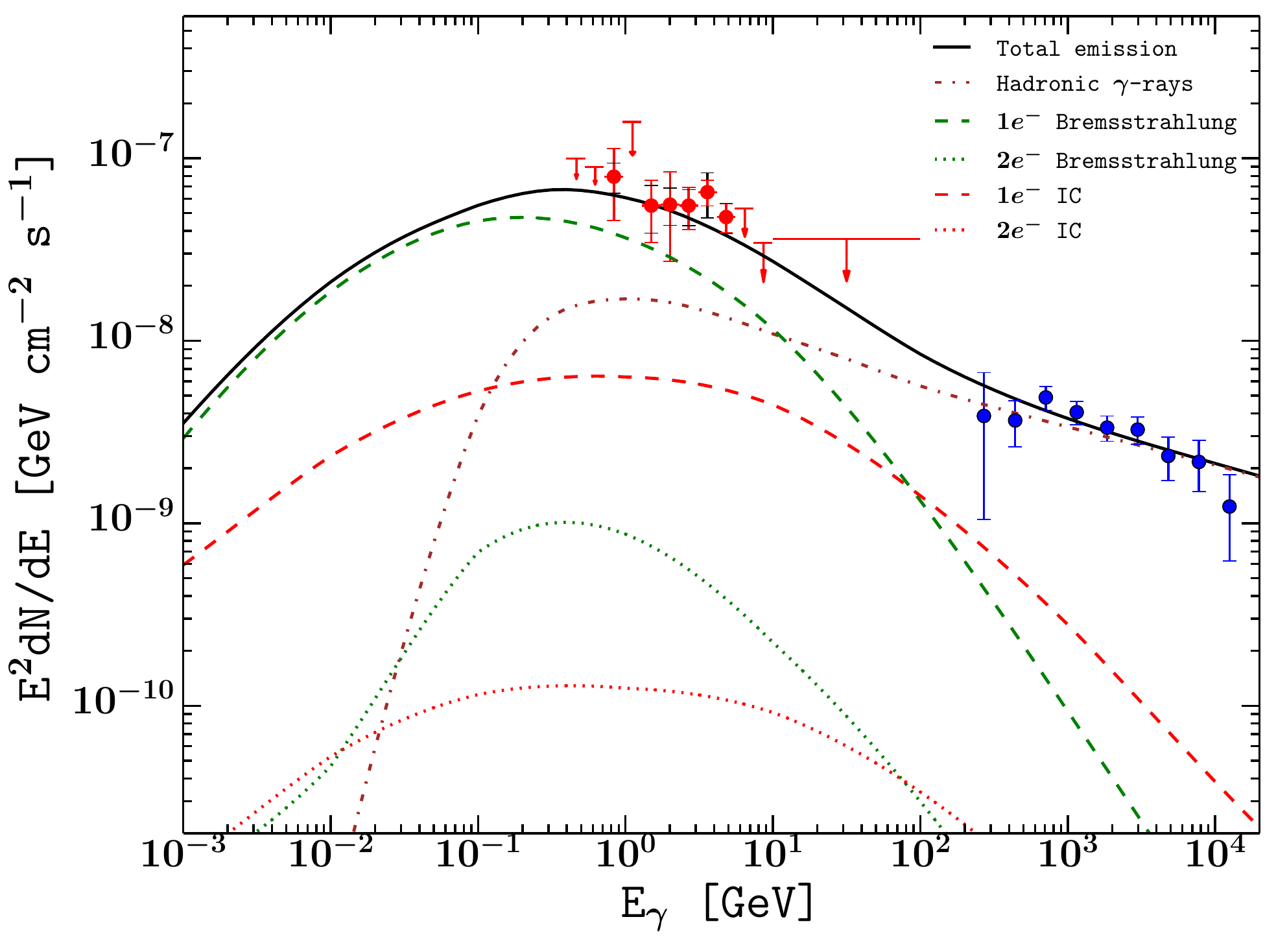}\\
\end{tabular}
\caption{ {\bf Broadband spectral energy distribution (SED) of the Ridge for the steady-state bremsstrahlung   solution when the gas density is fixed to $\overline{\langle n_H\rangle}=109.3\; {\rm cm^{-3}}$.}  \textit{Left Panel:} Radio flux density spectrum of the Ridge for the bremsstrahlung solution. Curves are : dashed black: primary electron synchrotron; short dashed (red) secondary electron synchrotron; solid (purple): primary + secondary synchrotron; dash dotted (black): primary + secondary synchrotron neglecting free-free absorption; solid with triangles (black): GSB; solid with triangles upside down (brown): free-free emission; and solid (black) total emission. Note that the lowest frequency (74 MHz) datum is interferometric, having been obtained with the VLA which is insensitive to the spatially slowly-varying  contribution of the GSB on the size scales of the Ridge solid angle. This is why the GSB (and its contribution to the total flux density  over the solid angle) is displayed with an unphysical cutoff at 100 MHz. }
\label{fig:plotfixed_nH}
\end{center}
\end{figure*}



%
%

\begin{figure*}
	\begin{center}
		\begin{tabular}{cc}
			
			\includegraphics[width=0.5\linewidth]{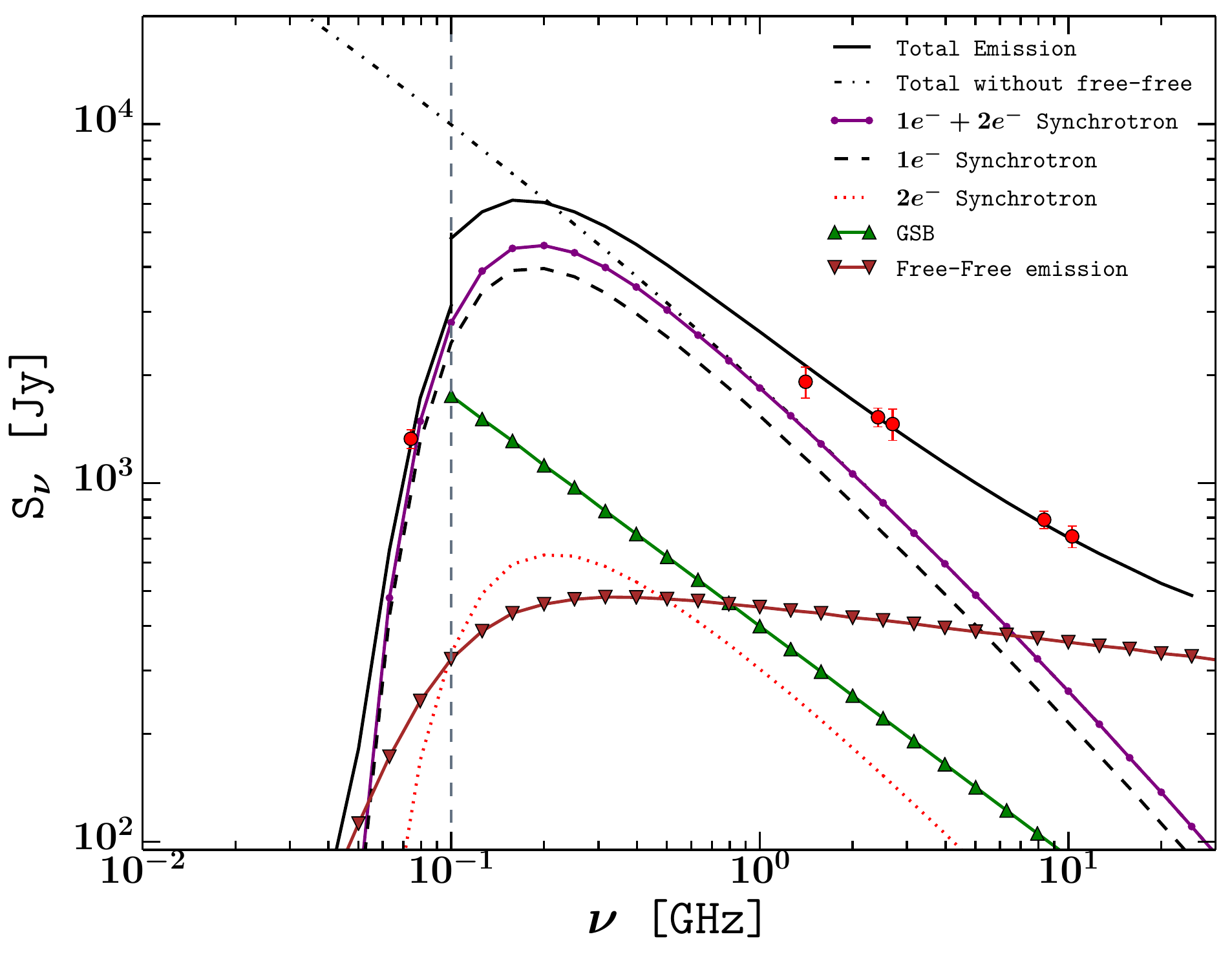} & \includegraphics[width=0.5\linewidth]{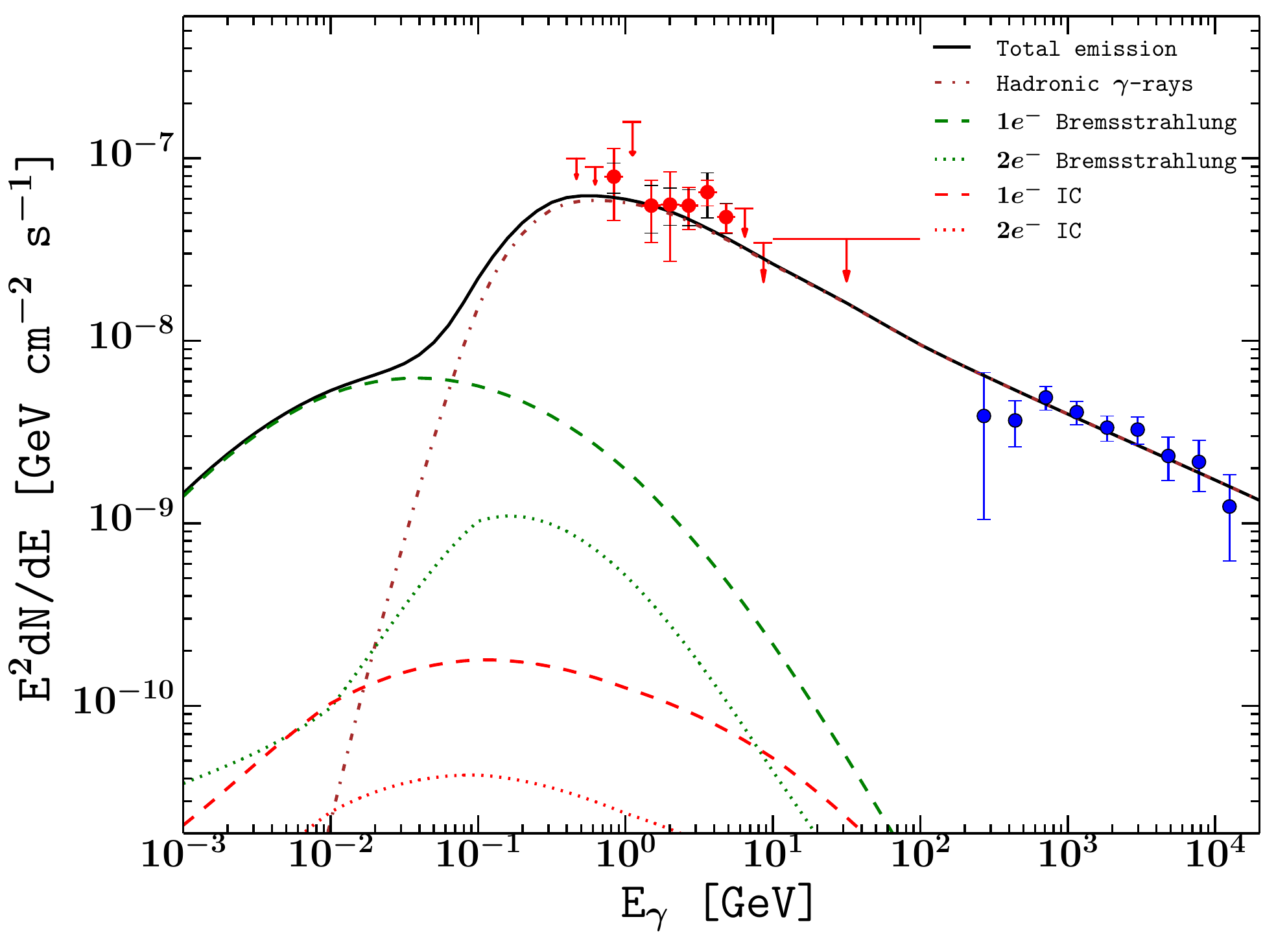}\\
		\end{tabular}
		\caption{{\bf Broadband SED of the Ridge for the steady-state $\pi^0$  solution when the gas density is fixed to $\overline{\langle n_H\rangle}=109.3\; {\rm cm^{-3}}$ and $\kappa_{\rm ep}=0.004$}. Curves are as given in the caption to fig.~\ref{fig:plotfixed_nH}. }
		\label{fig:plotfixed_nH_and_kappa}
	\end{center}
\end{figure*}

\subsection{\color{black}Non-steady-state fitting procedure}

\label{sec:nonssfit}

Our approach is to find the set of parameters that best describes the broad-band extended emission from the Ridge region. Specifically, we use a minimization procedure to determine the model parameters describing the injection spectrum of protons
and time of occurrence of the tentative supernovae explosions (or flare events from Sgr A$^{\star}$) giving rise to the CR population responsible for the extended emission from the region. Such a CR accelerator is assumed to be located at the Sgr A$^{\star}$ position in our analysis. 
{\color{Black} As we saw in the previous section, in the non-steady-state scenario,
	primary electrons cannot reach far
	enough in their loss times to account for the distributed 20 cm
	emission and also
	  we get a good fit to the Ridge with
	only a single non-thermal particle population (with secondary electrons
	explaining the observed synchrotron emission). For these reasons, we do not introduce an
	independent CR electron population.}
We set the diffusion coefficient parameters in Eq.~(\ref{eq:diffusionConstant}) to $\kappa=1$, which is  typical for cosmic-ray diffusion in the galactic disk \citep{Neronov},  and $\beta=0.3$ to have Kolmogorov diffusion. 
{\color{black} As can be seen from Eq.~(\ref{Rdiff}), the value of $\kappa$ is statistically degenerate with the flare age $t_0$.}
The required value of $R_{\rm diff}$ is determined by the  Ridge morphology. However, as this in turn is limited by the molecular-gas  distribution, the data only give a lower limit $R_{\rm diff}\gtrsim \tan\left(0.8^\circ\right)\times\; 8.5\; {\rm pc}=119$~pc. This implies our values for $t_0$ are lower limits, as a larger $t_0$ will not affect the spatial morphology. The lower limit on  $t_0$ can be changed by making the corresponding changes in $\kappa$ to preserve $R_{\rm diff}$. 
 \cite{Aharonian:2006au} found that CRs with a Gaussian morphology and a standard deviation of $0.8^\circ$ gave a good fit to the TeV data. We chose our flare ages so  that both our GeV and TeV flares resulted in CRs with a similar Gaussian distribution. This gave similar results to basing the flare age on setting $R_{\rm diff}=119$~pc for a typical energy range 
of the flare. 

In order to fit the radio data with our non-steady-state model, we also take into consideration thermal emission from the very complex environment of the Ridge. We assume a superposition of the thermal component with a spatially mixed non-thermal contribution given by Eq.~(\ref{eq:synchrotron}). Our calculations account for free-free absorption following the methods  described at length in \citet{Crocker2011b} and the supplementary material of~\citet{Crocker2010}. In short, the observed
flux density in the model is given by
\begin{equation}
F=\Delta \Omega\{S_{\rm NT}\exp[-\tau]+B[T](1-\exp[-\tau])+S_{\rm GSB}\},
\end{equation}
where $B(T)$ is Planck's function at the temperature $T$ and $\Delta \Omega$ is the Ridge solid angle.
The optical depth is parametrised as 
\begin{equation}
\label{eq:opticaldepth}
\tau(\nu) = \tau^{\rm ff}_0\left( \frac{\nu}{{\rm 0.325\;GHz}}\right)^{-2.1},
\end{equation} 
where $\tau^{\rm ff}_0$ is a parameter to be determined. Also, the Galactic synchrotron background (GSB) is modelled as
\begin{equation}
\label{eq:GSB}
S_{\rm GSB}={710\over \Delta \Omega} \left({\nu \over 408\; {\rm MHz}}\right)^{-\Gamma_{\rm GSB}}  [{\rm Jy}\, {\rm sr}^{-1}]
\end{equation} 
{\color{black} The fitting procedure to the radio data included an extra parameter related to the  contribution of thermal  emission. Namely, we calculate the
free-free flux density, $\Delta \Omega B[T](1-\exp[-\tau])$,
 at 10 GHz and allowed it to vary accordingly. This then determined the value of $\tau_0^{\rm ff}$. For the black body radiation spectrum we assume a kinetic temperature of 5000 K.
}
We also fit the non-steady-state model to the gamma-ray spectrum  in the GeV$-$TeV energy range. The full  fit is obtained via the numerical minimization algorithm \texttt{MINUIT}~\citep{minuit}. The global $\chi^2$ function entered to the minimization routine is given by
\begin{equation}
\label{globalchisq}
\chi^2 = \chi^2_{\rm radio} + \chi^2_{\rm GSB} + \chi^2_{\rm GeV} + \chi^2_{\rm TeV}\;,
\end{equation}
where the parameters over which we minimize the $\chi^2$ are: 
the free-free flux density at 10 GHz [Jy], $\Gamma_{\rm GSB}$ introduced in~\citet{Crocker2011b} ($ \chi^2_{\rm GSB}=\{\Gamma_{\rm GSB}-0.695\}^2/0.12^2$), 
the magnetic field $B$,  and the time of occurrence $t_0$ of the flare event. Also minimized over were  $K$ and $\Gamma$ which are the injection spectrum of protons (See Eq.~(\ref{eq:injectionSpectrum})). 
Notice however, that instead of the normalization $K$, we report the more convenient parameter the total emission energy, $E_{\rm total}  \equiv L_{\rm p} \Delta t$. 
Here $L_{\rm p}\equiv\int_{m_{\rm proton}}^\infty \Eg Q[\Eg]d\Eg$ is the injection luminosity {\color{black}(cf. eq.~\ref{eq:injectionSpectrum})}. In practice, we set the lower limit of the integral to 0, but as $Q$ is a power law in momentum space, for that range, it  makes a negligible difference if one rather has a power law in kinetic energy and integrates with a lower limit of $m_{\rm proton}$.

\section{Steady-state fits}

We also perform fits to the broadband spectrum of the Ridge in the steady-state limit.
We try to find self-consistent descriptions of the in situ, steady-state, non-thermal proton (for simplicity we neglect heavier ions) and electron populations in the Ridge and the environmental parameters describing the Ridge ISM such that the radiation from these populations reproduces the spectral data.
We assume that the non-thermal particles are injected into the Ridge ISM as power laws in momentum whereupon they undergo the energy loss and transport processes 
that shape their steady-state distributions.
Energy losses are the same 
as  the non-steady-state modelling: for CR protons,  ionisation at low energy and hadronic ($pp$) collisions at high energies;
for electrons,  losses are (from low to high energy) ionisation, bremsstrahlung, and synchrotron/inverse Compton emission (IC).

{\color{black}In steady state, if there is a significant energy dependence to the particle escape time (as for, e.g., the cosmic ray confinement time in the Galactic disk), $t_{\rm esc} = t_{\rm esc}[E_{\rm CR}]$, 
then the steady-state cosmic ray population in the region will have a spectrum given by $dN_{\rm CR}/dE_{\rm CR} \sim E^{-\Gamma_{\rm inj}} \ t_{\rm esc}[E_{\rm CR}]$.
Because $t_{esc}[E_{\rm CR}]$ is generally a decreasing function of energy,  energy-dependent transport leads to a softening with respect to the injection distribution in steady state
(as we have reviewed above, the non steady state diffusion case can, in contrast, actually lead to hardening with distance from the source).
For the GC system, we observe that the
spectra of the diffuse gamma-ray and radio continuum emission we are trying to model is rather hard,
implying that the distribution of parent cosmic rays emitting that radiation  is also hard,
consistent, in fact, with the spectrum  emerging directly from shock acceleration.
Thus we assume transport in our steady state modelling is energy-independent.
Broadly, this is
consistent with the evidence for the existence of a large-scale outflow from the region.
Note, however, one may also constrain, on energetics and other grounds, the total mass flux on such an outflow \citep{Crocker2012}.
In our single zone model, if the fitting returns a very large gas density where the radiation is occurring  (or such a gas density is imposed) and, at the same time, a small escape time, 
this could suggest an unphysically large mass flux out of the region ($> 0.5 \ \msun/$yr) {\it if} the energy-independent escape is affected by an outflow.
However, a high gas density plus short escape time need {\it not} be unphysical, however,  provided that we admit  the possibility that the effective diffusion coefficient be 
energy-independent (or have a mild energy dependence) {\color{Black}or that the cosmic rays stream out of the region at the Alfv\'en speed presumably 
 along the rather regular, poloidal magnetic field structure suggested by the large-scale non-thermal filament distribution \citep{Morris2006}.}
In such a case, cosmic rays diffuse out of the system quickly (and energy-independently) but the gas does not.
In any case, a further fitting parameter of our steady-state modelling is the energy-independent escape time, $t_\textrm{\tiny{esc}}$.
A sufficiently accurate approximation
(e.g.\ \citet{Crocker2011b} with  different notation)
 to the steady state distribution of cosmic-ray type $x$ (proton or electron) in the region accounting for both transport and losses is then 
\begin{equation}
\label{solutionSmpl}
{d n_x\over dE_x} \simeq \frac{t_\textrm{\tiny{loss}}\!(E_x) t_\textrm{\tiny{esc}}}{t_\textrm{\tiny{loss}}\!(E_x)+(\Gamma_\textrm{\tiny{inj}}-1)t_\textrm{\tiny{esc}}} { Q}_x\!(E_x)
\end{equation}
where $\Gamma_\textrm{\tiny{inj}}$ is the spectral index of the (assumed) power-law (in momentum) proton or electron spectrum at injection.

}

Radiation accompanying the loss processes listed above is self-consistently calculated to compare against the broadband spectral data on the region though note that we do not seek to reproduce morphological data with this fitting.
Hadronic collisions lead to the production of charged mesons (in addition to the neutral mesons whose decay is responsible for gamma-ray emission) 
generating final-state electrons and positrons; the radiation from these `secondary electrons' is self-consistently calculated in our steady state fitting, as for the non-steady state.
To calculate  IC cooling and emission, we assume an energy density for the interstellar radiation field of 90 eV cm$^{-3}$; \citep[see][for more details]{Crocker2012,Crocker2011b}.
The other parameters controlling cooling processes are the magnetic field amplitude (controlling synchrotron emission) and gas density (controlling ionisation, bremsstrahlung, and $pp$ losses); these parameters are left floating for the fitting process (though note  that, because of degeneracies, we end up fixing the gas density below to the volumetric average throughout the region).

Other floating parameters in the fit are: the spectral index of the GSB which contributes to the line-of-sight radio flux density over the Ridge's solid angle; the free-free (thermal bremsstrahlung) flux density at 10 GHz (from which we can self-consistently calculate the free-free $absorption$ at lower frequencies); the normalization at 1 TeV (in cm$^{-3}$ s$^{-1}$ eV$^{-1}$) and the spectral index, $\Gamma_p$, of the freshly-injected CR protons; and the normalization of the freshly-injected electrons at 1 TeV $relative$ to that of the protons ($\kappa_{\rm ep}$).
Note that we have explored allowing the electrons and protons' injection distributions to have independent spectral indices but, given the fitting prefers them to  be very similar, we set $\Gamma_e = \Gamma_p \equiv \Gamma_\textrm{\tiny{inj}}$ which is physically motivated by the expectation that non-thermal electron and proton populations in the region be accelerated on the same shocks.

Altogether there are 6 or 7 fitting parameters and 28 data points (6 radio data points, 12 Fermi points, 9 HESS points, and a data point which is the expectation for the spectral index of the GSB), implying 21 or 22 degrees of freedom unless otherwise noted.

\section{Results}
\label{sec:results}

\subsection{Non-steady-state Model}
\label{subsec:flaremodelresults}

The main results of the non-steady-state model are shown in Fig.~\ref{fig:spectralfitBlackHole} and Fig.~\ref{fig:spatialfitBlackHole}, where we reproduce the observed broadband radiation spectrum as well as the spatial gamma-ray distribution (at GeV and TeV energies) with two flares from the central source.  It is interesting to note that, very likely, there have been a series of flares with different energetic properties occurring throughout the lifetime of Sgr A$^{\star}$. Our fit preferred a model in which the impulsive events tentatively occurred $3\times 10^{5}$ and $2\times 10^{4}$ years ago for the GeV and TeV flares respectively.  The total energy required to inject relativistic protons capable of accounting for the extended radiation from the region were $2\times 10^{52}$ and $2\times 10^{50}$ erg  each. 
 The
latter is a reasonable match for a single supernovae remnant, the former is not,
so presumably would require a burst event from the super-massive black hole.
The duration of both flares was chosen to be $10$ years, however, 
as long as the flare duration is much less than the flare age ($t_0$), only the total injected energy affects the predicted gamma-ray spectrum.
Details of our best-fit parameters are provided in Table~\ref{tab:FitGeVTeVFlareModel}.

In  Fig.~\ref{fig:spectralfitBlackHole} we show that it is possible to fit the entire gamma-ray domain with hadronic photons resulting from the scattering of protons with hydrogen gas in giant molecular clouds. The same interaction process produces charged mesons (mainly $\pi^{\pm}$) whose subsequent decay creates a non-thermal population of relativistic electrons and positrons. The synchrotron light emitted by such particles, in conjunction with thermal emission from the region, give an acceptable fit to the data over the radio band.

\begin{table*}

\begin{tabular}{lrr}
\hline\hline
\centering
Model & $2\log (\mathcal{L}/\mathcal{L}_{\rm base})$ & {\rm dof}$_{\rm base}-${\rm dof}\\  \hline 
{\color{black} Base=2FGL$+$``bgkA''$+$``New point source'' $-$``the Arc''$-$Sgr B+Spherically symmetric source}&  0 & 0  \\ \hline
Base$+$
20 cm template & 176  & 3 \\ 
Base$+$
HESS residual template   &   149  & 3\\
Base$+$
Best-fitting two-flare model template   & 194    &6 \\ \hline\hline
\end{tabular}

\caption{\label{tab:LogLikelihoods} The likelihoods evaluated in compiling the above table are maximized with a broad band analysis using the Fermi Tools. Alternatives models of the GC in the 200 MeV$-$100 GeV energy range are listed.
Each point source in the model has degrees of freedom ({\rm dof}) from its spectrum and two extra {\rm dof} from its location. The spectra for the Ridge templates are modeled by a broken power law, except for our two-flare model template, where we use the deduced gamma-ray spectrum. In our two-flare model, the relevant {\rm dof} are the two flare-ages and the two sets of injection spectrum parameters. While the spectra for the ``spherically symmetric source'' templates are modeled by a log parabola which has enough flexibility to mimic a good fitting dark matter or unresolved millisecond pulsars spectra \citep{gordonmacias2013}.
 }
\end{table*}

\begin{table*}
	{\color{black}
		\begin{tabular}{|p{0.78 in}|p{0.35 in}|p{0.23 in}|p{0.2 in}|p{0.85 in}|p{0.3 in}|p{0.5 in}|p{0.3 in}|p{0.5 in}|p{0.47 in}|p{0.47 in}|p{0.42 in}|}
			\hline\hline
			Steady-State Model & $B$ $\left[ \mu {\rm G}\right]$ & $\Gamma_{\rm e},\Gamma_{\rm p}$& $\Gamma_{\rm GSB}$&  Normalization of protons at 1 TeV  $\left[{\rm cm^{-3}\;s^{-1}\; eV^{-1} }\right]$& $\kappa_{\rm ep}$& $t_{esc}$ $\left[\rm years\right]$ & $\overline{\langle n_H \rangle}$  $\left[\rm cm^{-3}\right]$ & free-free flux density at 10 GHz $\left[\rm Jy \right]$ & $L_{\rm p}$ $\left[{\rm erg\; s^{-1}}\right]$ & $L_{\rm e}$ $\left[{\rm erg\; s^{-1}}\right]$  & $\chi_{\rm min}^2/{\rm dof}$  \\  \hline 
					Bremsstrahlung solution with $\overline{\langle n_H \rangle}$ fixed & $140 $ & 2.36 & 0.7 & $2.3 \times 10^{-38}$ & 0.14 & $9 \times 10^4$ & 109.3 & 330 & $10 \times 10^{37}$ & $3.7 \times 10^{37}$ &$ 10.3/21$ \\ \hline
			$\pi^0$-solution with $\overline{\langle n_H \rangle}$ and $\kappa_{\rm ep}^{\tiny{Bell}}$ fixed& 500 & 2.47 & 0.6 & $11 \times 10^{-38}$ & 0.004 & $3 \times 10^4$ &109.3 & 360 & $7.5 \times 10^{38}$ & $ 1.1\times 10^{37}$& $12.5/22$   \\ \hline \hline
		\end{tabular}
		
		\vspace{0.2cm}
		\caption{\label{tab:crockerresults} Best fit parameter values obtained for two different solutions of the  steady-state model for the Ridge region. }
	}
\end{table*}

\begin{table*}
	{\color{black}
		\begin{tabular}{llllp{0.8 in}llp{0.7 in}}
			\hline\hline
			Steady-State Model & $B$ $\left[ \mu {\rm G}\right]$ & $\Gamma_{\rm e},\Gamma_{\rm p}$& $\Gamma_{\rm GSB}$&  Normalization of protons at 1 TeV  $\left[  {\rm cm^{-3}\;s^{-1}\; eV^{-1} }\right]$& $\kappa_{\rm ep}$& $t_{esc}$ $\left[\rm years\right]$ & free-free flux density at 10 GHz $\left[\rm Jy \right]$ 
			 \\  \hline 
			Bremsstrahlung solution with $\overline{\langle n_H \rangle}$ fixed & $130\pm 20$ &$ 2.34_{-0.07}^{+0.06}$ & $0.7\pm0.1$& $(2\pm 1)\times10^{-38}$ &$ 0.2\pm 0.1$ & $(9\pm 6)\times10^4$  & $320\pm 20$ 
			\\ \hline
			$\pi^0$-solution with $\overline{\langle n_H \rangle}$ and $\kappa_{\rm ep}^{\tiny{Bell}}$ fixed& $490\pm80$ & $2.47\pm 0.02$ & $0.6\pm0.1$ & $(11\pm 7)\times10^{-38}$ & 0.004 & $(3\pm 2)\times10^4$ & $360\pm 20$
			 \\ \hline \hline
		\end{tabular}
		\vspace{0.2cm}
		\caption{\label{tab:crockerresultsmean}
		Mean parameter values obtained from marginal distributions assuming a non-negative prior for all parameters. The errors correspond to the 68\% confidence intervals.
			 }
	}
\end{table*}
\begin{figure}
\centering
\includegraphics[width=1.0\linewidth]{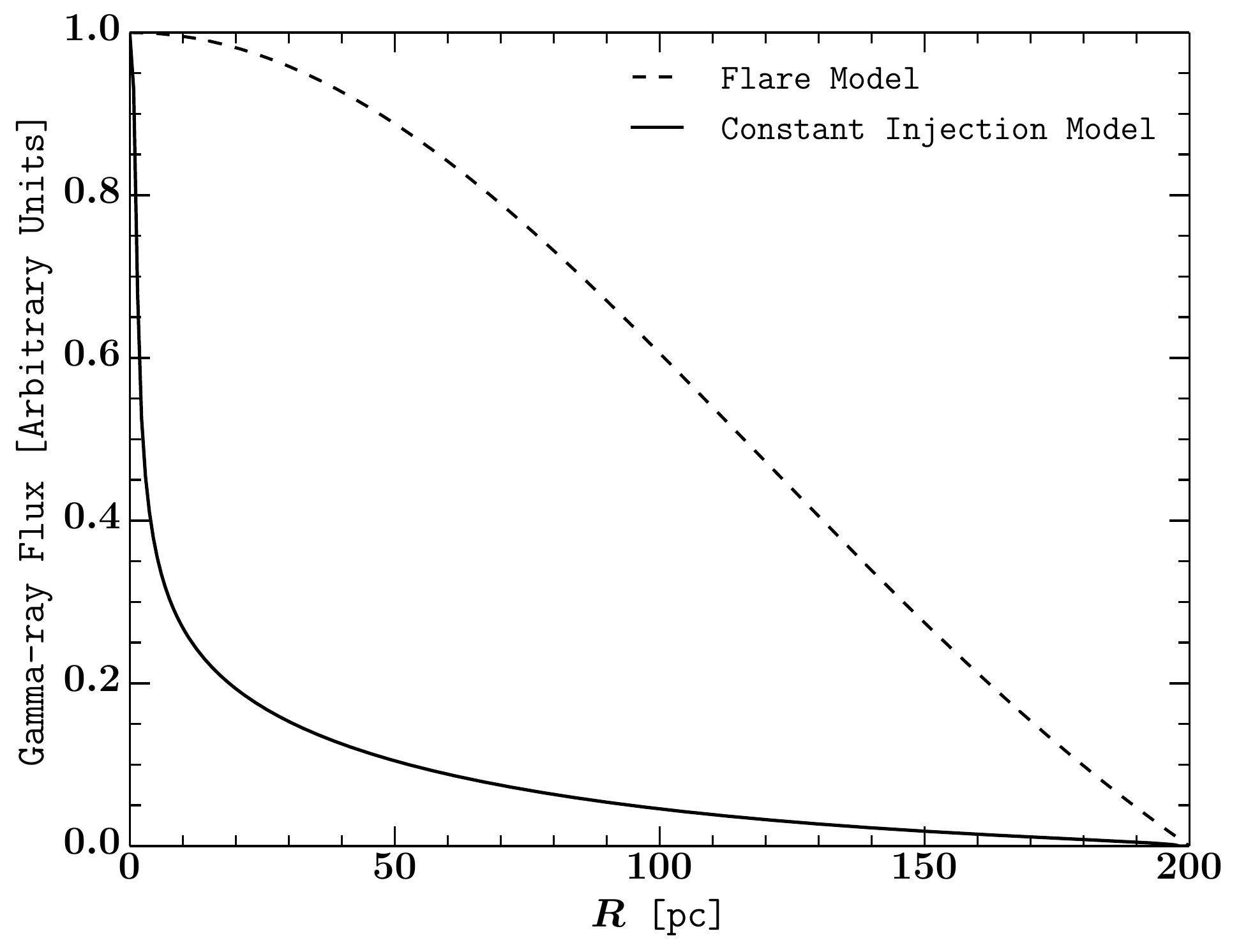} 
\caption{ \label{fig:comparison} Normalized  {\color{black} line of sight integrated gamma-ray flux profiles of the GC ($\abs{ l } < 0.8^{\circ}$ and $\abs{ b }<0.3^{\circ}$) in the 270 GeV$-$12.5 TeV energy range. The dashed line  corresponds to the
	\citet{Aharonian:2006au} proposed
	 flare which occurred $\sim 10^4$ years ago. The solid line illustrates the distribution obtained when the best-fit parameters proposed in~\citet{Aharonian:2006au} are used in a  $ 10^4$ year constant emission scenario.    }}   
\end{figure}

As discussed in Sec.~\ref{sec:nonssfit}, the parameters of our model are also constrained by additional information extracted from the spatial morphology of the gamma rays  in the Ridge.  Although we used the dimensions of the Ridge to set the flare times, in principle the likelihood analysis could include a fit to the spatial morpholology. For the GeV data, we will do this rigorously in Sec.~\ref{subsec:gevspatial} and see that it matches our considerations in Sec.~\ref{sec:nonssfit}. For the TeV data, higher resolution gas maps were utilized in~\citet{Aharonian:2006au} and we have used the constraints there in our fitting procedure describe in Sec.~\ref{sec:nonssfit}. In this subsection we compare the brightness profiles of the data and model predictions and show that the general trend of data is matched. 

In the top panel of Fig.~\ref{fig:spatialfitBlackHole} we show a two-dimensional representation of our best-fitting gamma-ray spatial distribution.
In the bottom panel of Fig.~\ref{fig:spatialfitBlackHole}, 
predictions of our model are shown with blue histograms. The red dashed histograms show the averaged gas distribution obtained from maps provided by~\citet{Ferriere}, while red continuous line are taken from CS line emission observations~\citep{Aharonian:2006au}. Visual inspection shows that gas maps in~\citet{Ferriere} are consistent with the one used by~\citet{Aharonian:2006au} but of lower resolution. The fact that maps in~\citet{Ferriere} are coarser than the CS maps, explains why our model fails to account for the dip at $l\simeq +0.3^{\circ}$ in bottom-right panel. The same argument can be applied to explain a deficit in gamma rays at $l\simeq +1.5^{\circ}$ in bottom-left panel of Fig.~\ref{fig:spatialfitBlackHole}. We thus note that a more detailed gas map of the Ridge  will very likely improve to an acceptable range the quality of the spatial fit for the non-steady-state model.

As a consistency check, we also evaluated the TS value of the GeV spatial map obtained from our non-steady-state model predictions using the Fermi-Tools software package~\footnote{\url{http://fermi.gsfc.nasa.gov/ssc/data/analysis/documentation/}}. These results are shown in Table~\ref{tab:LogLikelihoods}.

\subsection{Steady State Results}
{\color{black} We find two very satisfactory parameter regimes that can be broadly described as: 

(i) $\sim$GeV dominantly primary electron bremsstrahlung, $\sim$TeV dominantly hadronic emission, and radio dominantly primary electron synchrotron, with a large gas density; 

and 

(ii)  $\sim$GeV $and$ $\sim$TeV dominantly hadronic emission, and radio dominantly primary electron synchrotron with high gas density and slow escape (approaching thick target limit).

Each of these is described in more detail and illustrated below.
We remark in passing that we have also found a class of solutions with $\sim$GeV $and$ $\sim$TeV dominantly hadronic emission and  radio dominantly $secondary$ electron synchrotron
but these require very strong magnetic fields and are somewhat statistically disfavoured by the fitting procedure so we do not pursue them further here. Also, there are $\sim$GeV $and$ $\sim$TeV dominantly hadronic emission, and radio dominantly primary electron synchrotron with low gas density and fast escape (consistent with a  wind) solutions. However, due to large degeneracies between the gas density and other parameters, in this article we just consider the gas density giving by the ~\citep{Ferriere} of $\overline{\langle n_H \rangle} = 109.3$ cm$^{-3}$ for the Ridge region for both the steady-state and non-steady-state cases.}

\begin{figure*}
\begin{center}
\begin{tabular}{cc}

\includegraphics[width=0.5\linewidth]{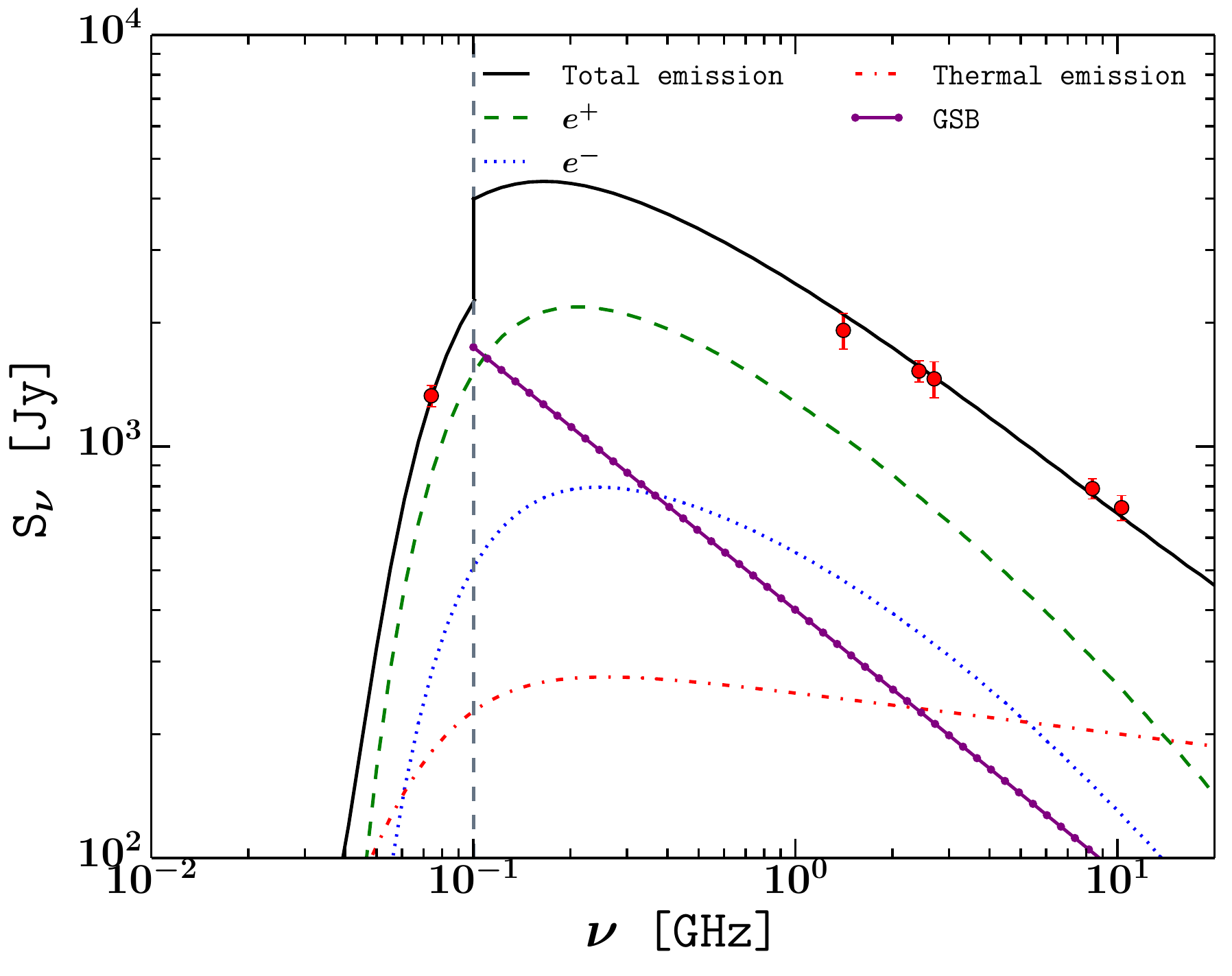} & \includegraphics[width=0.5\linewidth]{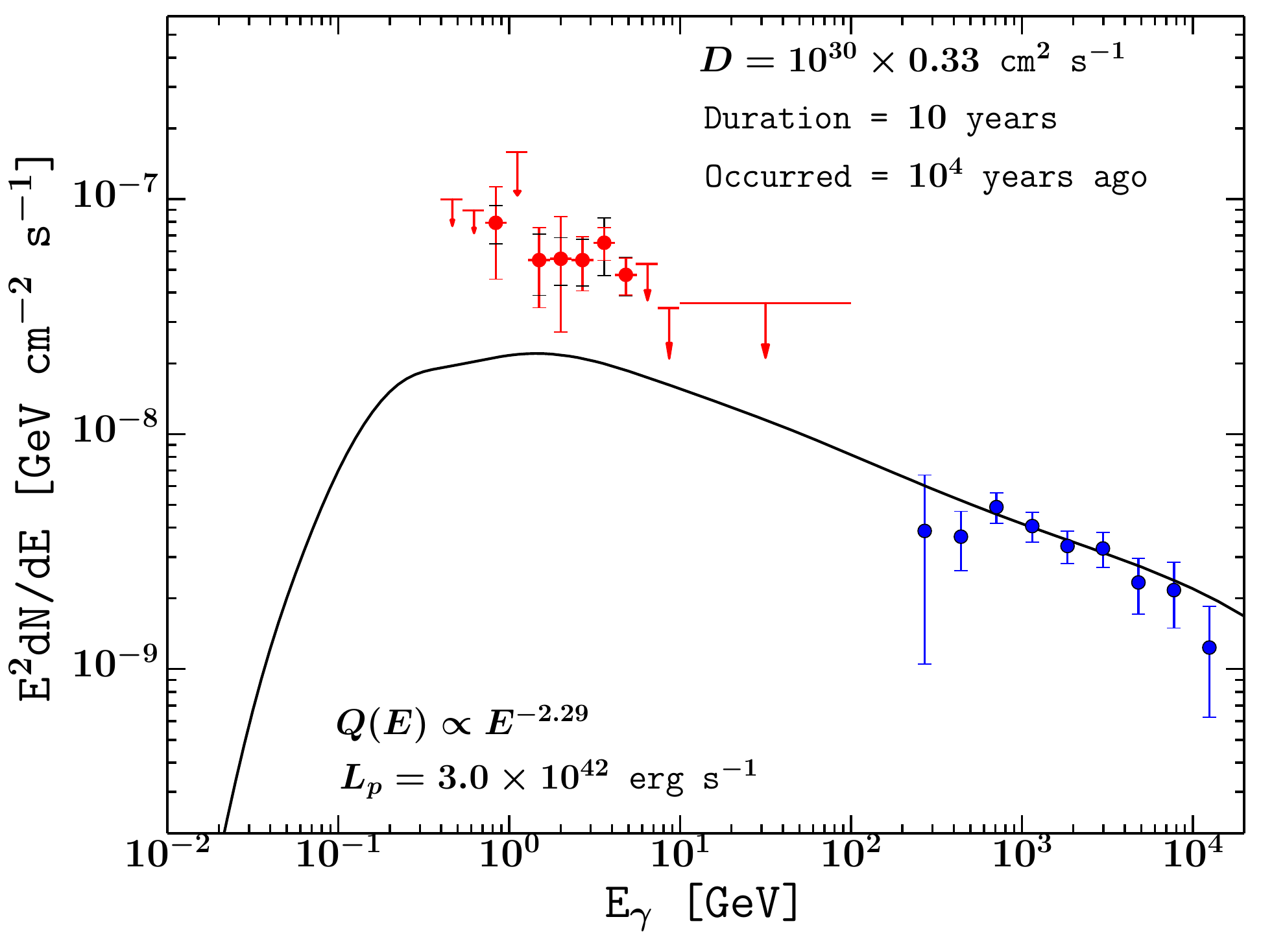}\\
\end{tabular}
\caption{{\bf Flare model proposed by the HESS collaboration}~\citep{Aharonian:2006au}: The model consists of a single flare that might have occurred $\sim 10^4$ years ago and lasted for approximately $10$ years. We use the gas maps provided in~\citet{Ferriere}. Note that the HESS team assumed an energy independent diffusion coefficient. Radio data is well fitted by synchrotron emission resulting from secondary CR electrons and positrons. For this case, the best fit magnetic field amplitude was $B=495$ $\mu$G.  }
\label{fig:HESSsolution}
\end{center}
\end{figure*}

\subsubsection{GeV bremsstrahlung solution}

{\color{black} This solution (See Fig.~\ref{fig:plotfixed_nH}, Table~\ref{tab:crockerresults}, and Table~\ref{tab:crockerresultsmean}), for which $\chi_{min}^2/{\rm dof}$ = 10.3/21, is similar to that previously found by \citet{Yusef-Zadeh2013}.

Consistent with previous work \citep{Crocker2010} we find this  fit  (and the others) prefers a strong magnetic field in the 100 $\mu$G range, specifically $130\pm20$ $\mu$G for this case.

Note that the energy-independent escape here {\it cannot} be attributed to a wind  \citep[for the escape time determined here and volumetric average gas density, the wind mass flux would correspond to an unphysically  value much larger than  $1 \msun$/year:][]{Crocker2012}. Therefore, the energy-independent escape would need to be due to energy-independent diffusion or streaming along the regular magnetic field at the Alfven velocity.

\subsubsection{$\pi^0$ with $\kappa_{\rm ep}^{\tiny{Bell}}$  and $\overline{\langle n_H \rangle}$ solution}
For the bremmstrahlung case given above we find a very large power going into freshly-accelerated electrons relative to that going into protons.
Such a regime is quite different to that usually inferred from individual supernova remnants or the Galaxy-at-large \citep[e.g.,][]{Thompson2006,Thompson2007} where $L_{\rm e} \lsim 0.1 \ L_{\rm p}$ (where $L_x$ is the luminosity going into freshly accelerated particles of type $x$).
If physical, such a large electron power must be connected to the unusual conditions in the GC environment.
For instance, the requisite electron acceleration might be associated with magnetic field reconnection occurring in the  non-thermal radio filaments \citep{Yusef-Zadeh1984} found uniquely in this region.
If this is the case, however, it remains unexplained why the fitting prefers $\Gamma_e \simeq \Gamma_p$ with a value for the injection spectral index typical with expectation for first-order Fermi acceleration at astrophysical shocks.

We have also explored whether restricting $\kappa_{\rm ep}$ to the expectation for first-order shock acceleration
\citep{Bell1978} so that it satisfies $\kappa_{\rm ep} \equiv (m_p/m_e)^\frac{1-\Gamma_\textrm{\tiny{inj}}}{2} 
\equiv \kappa_{\rm ep}^{\tiny{Bell}}$.
Again we find an acceptable fit to the data.
%
See Fig.~\ref{fig:plotfixed_nH_and_kappa} and Table~\ref{tab:crockerresults}.   
Note that here also the energy-independent escape  {\it cannot} be attributed to a wind because, again, for the escape time determined here and volumetric average gas density, the wind mass flux would correspond to   $\gg 1 \msun$/year, \citep{Crocker2012}  but must be due to energy-independent diffusion or streaming.}

\section{Discussion}

\begin{figure}
	\centerline{\includegraphics[width=0.5\textwidth]{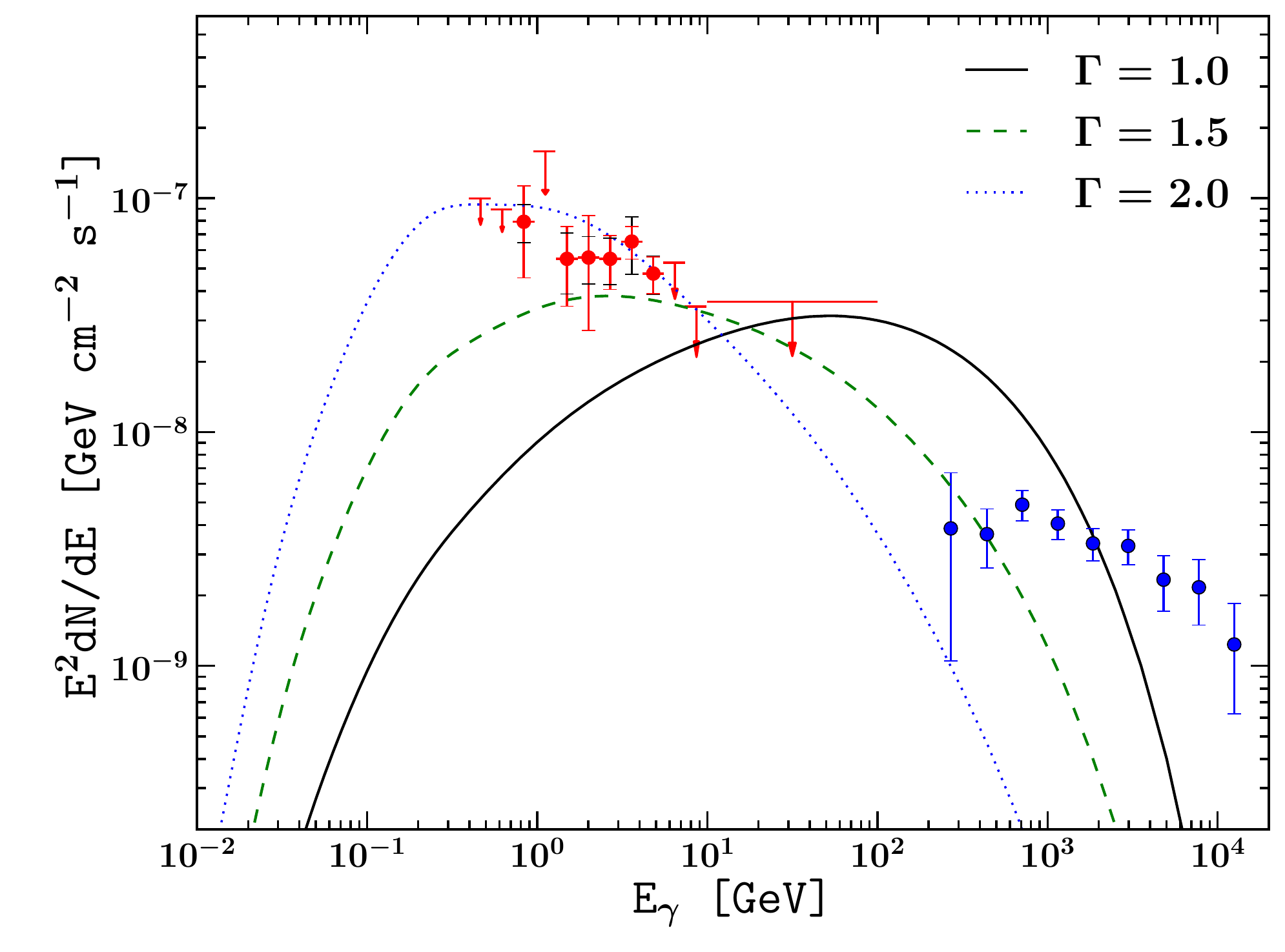}}
	\caption{ Impact on the SED of the $3\times 10^5$ year old flare for different choices of the spectral slope $\Gamma$ of the injection spectrum $Q(E)$. See Sec.~\ref{sub:singlevstwoflares} for details.}
	\label{fig:why_one_flare_only}
	
\end{figure}

\begin{figure*}
\begin{center}
\begin{tabular}{c}
\includegraphics[width=\textwidth]{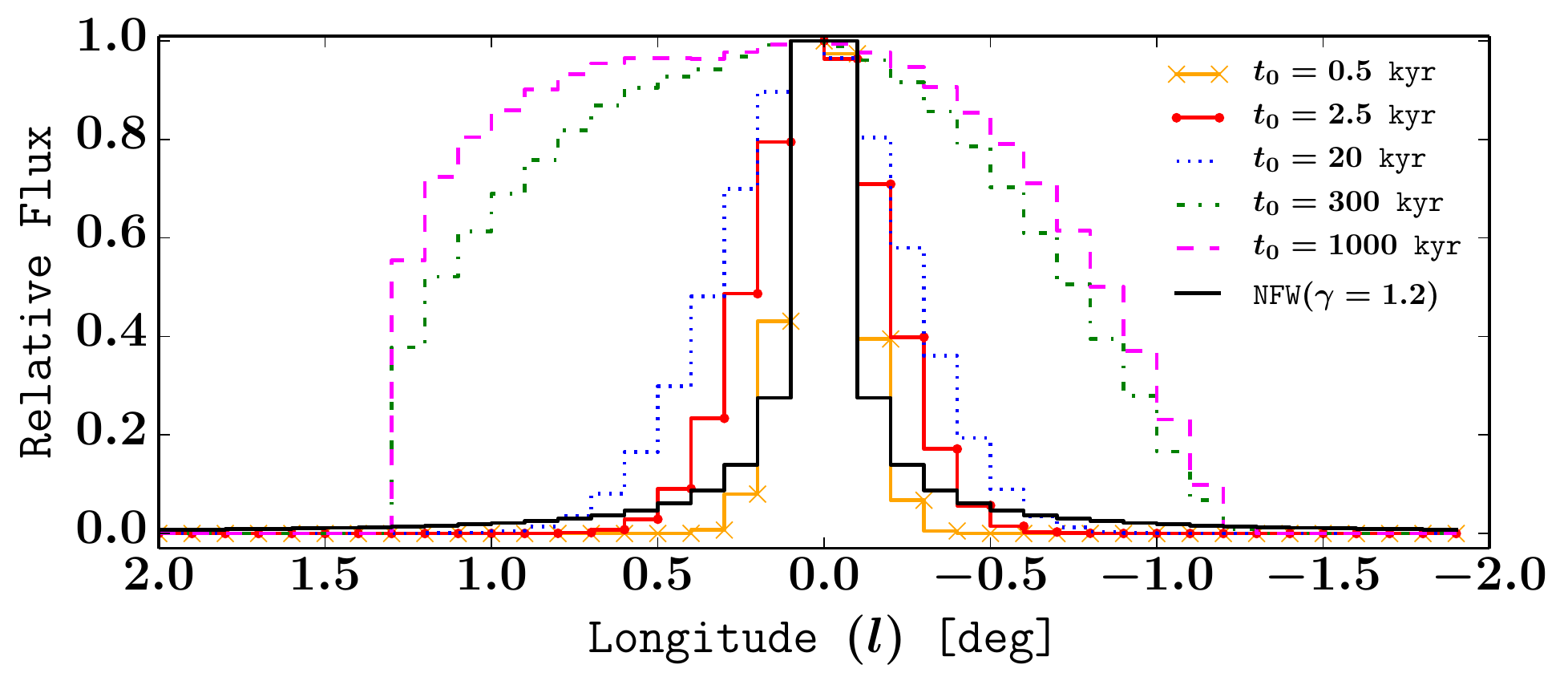}\\
\includegraphics[width=\textwidth]{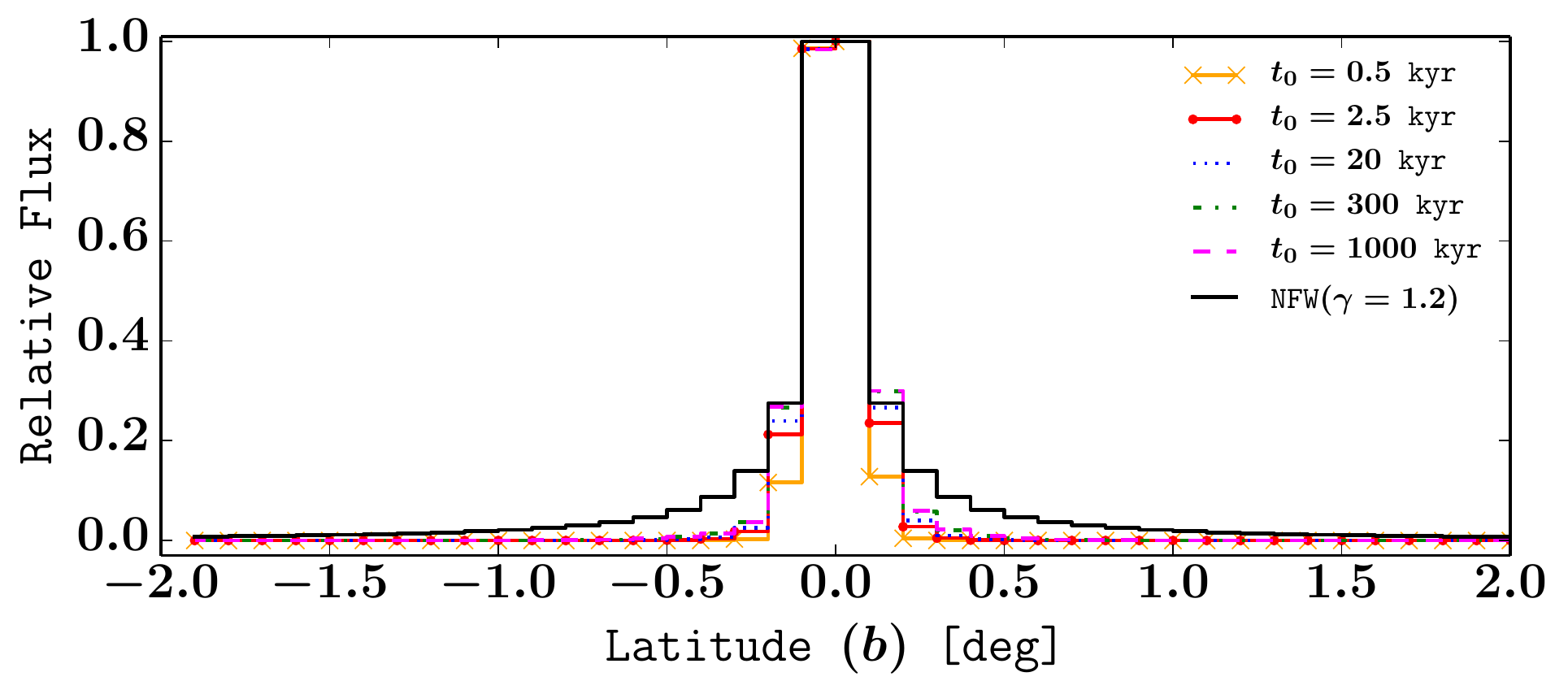}  
\end{tabular}

\end{center}
\caption{ Brightness profile of the Ridge in $0.2-100.0$ GeV energy range for the flare model with a Kolmogorov spectrum ($\kappa=1.0$ and $\beta=0.3$). The injection spectrum is described by an exponential cut-off in momentum with slope $\Gamma=2.0$. We display burst events of increasing age for which the fluxes are normalized to the maximum. 
For comparison purposes we also include a generalized NFW profile, with inner slope $\gamma=1.2$, which provides a good fit to the spherical excess.
}
\label{fig:SpatialMorphologyFlareModel}

\end{figure*}

In this section, we  discuss four different proposals for the Ridge gamma-ray data: A flare event from the GC with energy independent diffusion {\color{black}(or, more generally, transport)}, two flare events  from the GC with Kolmogorov diffusion, a continuous emission from the GC, and a steady-state model with cosmic ray injection 
occurring throughout the Ridge with energy-independent transport.

We used the formalism of~\citet{Aloisio}, which was recently utilized by~\citet{Chernyakova:2011zz,LindenProfumo2012,LindenLovegroveProfumo2012} in the point-source analysis of HESS J17452$-$290. Indeed, if we compute the expected cosmic-ray density distribution from the Ridge using similar parameters to those obtained in~\citet{Aharonian:2006au} but assuming constant injection  of relativistic protons (instead of the flare emission used in~\citet{Aharonian:2006au}), we find a bad fit to the diffuse gamma-ray data. 
As can be seen in Fig.~\ref{fig:comparison}, a continuous emission from the GC produces a more peaked spatial profile
compared to a flare event.

In this sense, a constant injection model should be better suited to explain the point source nature of Sgr A$^{\star}$~\citep{Chernyakova:2011zz,LindenLovegroveProfumo2012}.  An excessively peaked profile, for the HESS Ridge,  from the GC was also found for a constant source model by \citet{Melia}, who used an approach based on simulating the trajectories of individual protons.   However, as discussed above, our analysis does show that if a flare model is considered, a single CR accelerator located in the Galaxy Center can consistently explain the broad band diffuse photon data from the Ridge.  

Notice that our flare models neglect diffusive reacceleration throughout the Ridge. The time
scale of this acceleration is \citep{amano2011} $t_{acc}=D /v_A^2,$ where $D$ is the spatial
diffusion coefficient and $v_A$ is the Alfv\'en speed. For our  two-flare model best fit value of $B$, we have $v_A=B/\sqrt{4\pi\rho}\approx40$ km/s which gives  $t_{acc}\sim 10^7$ years for $E\sim 1$
GeV. 
This justifies neglecting reacceleration, in our case, as our GeV producing flare happened $3\times10^5$ years ago. A similar calculation for our TeV flare also justifies neglecting  reacceleration  throughout the Ridge.

{\color{black}
	We used the radio data compilation from \cite{Crocker2011b} rather than the radio data from \cite{Yusef-Zadeh2013}. The difference  between the two is that the latter  presents background subtracted fluxes. We prefer to marginalize over the background so as to account for the additional uncertainty.
	The data in \cite{Yusef-Zadeh2013} also do not  include the 5\% systematic error associated with absolute
	calibration, although this could easily be added.
	
	The p-values of our global $\chi^2$ of Eq.~(\ref{globalchisq})  are on the high side, but generally within acceptable bounds with our  highest one being for the Bremsstrahlung solution with $\overline{\langle n_H \rangle}$ fixed case in Table~\ref{tab:crockerresultsmean} which had a  p-value of 0.97. 
	 This may indicate that the radio data error bars have been overestimated. 
}

\subsection{Single flare versus multiple flares model}

\label{sub:singlevstwoflares}

In \cite{Aharonian:2006au} the $\sim$TeV diffuse spectrum and spatial gamma-ray distribution was explained with a single impulsive injection of CR protons occurring near the dynamical center of the Milky Way Galaxy. Their fit preferred a set of parameters for which the CR density is described by a Gaussian distribution (centered on Sgr A$^{\star}$) with a dependence on distance to the GC given by the standard deviation $\sigma=0.8^{\circ}$. They estimated that a central source of age $\sim$10 kyr and energy independent diffusion coefficient $D=\eta \times 10^{30}$ cm$^2$ s$^{-1}$, where $\eta \leq 1.0$, can account for the observed morphology and spectrum of the TeV-gamma rays. In view of more recent measurements of diffuse radio emission~\citep{Crocker2011b,Yusef-Zadeh2013} and GeV-gamma rays~\citep{Yusef-Zadeh2013,MaciasGordon2014} from the same region, it is not immediately obvious whether this model would still be adequate at a multiwavelength level.

{\color{black}We reproduced the results reported~\citet{Aharonian:2006au} and evaluated the goodness of fit for the HESS, Fermi-LAT and radio data.
We fixed  the injection spectral index to the value used in \citet{Aharonian:2006au}, which had $\Gamma=2.29$. This model, which is plotted in Fig.~\ref{fig:HESSsolution}, has $\chi_{min}^2/{\rm dof}=23.3/20$
which corresponds to a 	p-value$>0.001$. However, as can be seen it is consistently below the data points in the right panel of Fig.~\ref{fig:HESSsolution} and the fit has a Durbin-Watson statistic which has a p-value considerably below 0.001.


As this single flare has an energy independent diffusion coefficient and  the escape time is less than the energy loss time (see Fig.~\ref{fig:tploss}), we expect to recover a steady state solution. As was seen in Fig.~\ref{fig:plotfixed_nH_and_kappa} and the last row of Table~\ref{tab:crockerresultsmean}, if the injection spectral index is 2.47, then a good fit to the TeV and GeV data set can be obtained. We have confirmed that a single flare, occuring $10^4$ years ago, with $\Gamma=2.47$, and $D=0.33\times 10^{30}$ cm$^2$s$^{-1}$ reproduces the hadronic component in Fig.~\ref{fig:plotfixed_nH_and_kappa}.
}

For the single flare model,  the CR spatial density distribution in the $\sim 0.2-100.0$ GeV and $0.2-12.5$ TeV energy bands is the same  because the diffusion coefficient is taken to be energy independent. In principle, the energy losses are energy dependent, but as can be seen from Fig.~\ref{fig:tploss}, the energy-loss time scale is much greater than the 
$t_0=10^4$ years used in this case. Therefore, energy losses are not a significant factor for our single flare fit.

Notice that the diffusion coefficient may well be energy dependent, as for example in the case of a Kolmogorov spectrum of turbulence. 
The impact of assuming a Kolmogorov energy-dependent diffusion $D(E)$ on the resulting gamma-ray spectra is demonstrated in Fig.~\ref{fig:spectralfitBlackHole}. The most salient feature is that, for this case, a single CR-injection event fails to explain all the observations. We are thus required to invoke a model consisting in the superposition of two flares. 

It is instructive to understand why a single impulsive event cannot be accommodated to the diffuse Ridge gamma rays when the diffusion coefficient is a function of energy: For the best fit parameters in Table~\ref{tab:FitGeVTeVFlareModel} our $2\times 10^{4}$ year old TeV flare cannot have a strong impact on the observed GeV gamma-ray distribution, since most of the $\sim$GeV protons from the flare are still trapped in the surroundings of the central source, i.e. $R_{\rm diff}=\sqrt{2\lambda({\rm 1\ GeV, 2\times 10^4 \ years})} \approx 50$ pc.  On the other hand, the $3\times 10^5$ years old GeV flare cannot explain the TeV data because most of the very-high-energy protons have left the region. Fig.~\ref{fig:why_one_flare_only} illustrates how even if we harden sufficiently the injection spectrum of protons, the fit cannot be ameliorated. This indicates that we need a fresher injection of protons to explain the TeV-gamma rays.

\subsection{Resolving the spatial extension of the flare model}
\label{subsec:gevspatial}

As can be seen from  Fig.~\ref{fig:SpatialMorphologyFlareModel}, a  $t_0=0.5$ kyr or younger flare results in a relatively spherical morphology. This results from the CRs not yet reaching the CMZ boundary.
  {\color{black} We now consider the prospect that a  flare could explain the
 GC spherically symmetric extended source in addition to the Ridge.}
 
The spatial morphology resulting from a flare depends primarily on the flare age and the diffusion coefficient. To test the spatial morphology of the flare independently of the flare spectrum, we model  the flare spectrum with a flexible broken power-law of the form 
\begin{equation}
  \frac{dN}{dE}=N_0 \times 
  \begin{cases}
    \left(\frac{E}{E_b}\right)^{-\Gamma_1} &\text{if $E<E_b$} \\
\left(\frac{E}{E_b}\right)^{-\Gamma_2} &\text{otherwise.}
  \end{cases}
   \label{eq:BrokenPl}
\end{equation} where $N_0$ is a normalization constant and $E_b$ is the break energy. This could always be generated provided sufficient flexibility was allowed in the flare injection spectrum.

{\color{black} To test whether the flare model can explain the spherical excess, we include both a flare model and a well fitting spherical excess model. We check if the inclusion of the flare model is able to decrease the significance of the spherical excess model.}
 The results of this analysis are shown in Fig.~\ref{fig:TSvaluesFlareModels}. A full Fermi-tools analysis was done for each flare age. Both the flare template and the GC spherically symmetric extended source template were included. {\color{black} As can be seen, there is no choice of flare age which non-negligibly affects the significance of the spherically extended source.} As can be seen from Fig.~\ref{fig:SpatialMorphologyFlareModel}, this inability of our flare models to describe the spherically symmetric extended source is due to a mismatch in the radial fall-off and extension of the younger flares ($t_0\leq 0.5$ kyr) relative to the spherically symmetric source which requires
a fall off of flux like $r^{-2.4}$ out to at least $1^\circ$  \citep{gordonmacias2013},
while the older flares under consideration produce an excessively ridge-like morphology. The improvement in the flare fit levels off at $t_0\sim3\times 10^5$ yr as then the CRs have reached the edges of the CMZ.  As can be seen from Table.~\ref{tab:LogLikelihoods}, the best fit flare model provides just as good a fit as the HESS or 20 cm templates. { \color{black} So although the flare model is unable to describe the GC spherical excess it does provide a good description of the Ridge excess.}

{\color{black}\subsection{Steady-state model}}
\begin{figure}
\begin{center}
\includegraphics[width=0.5\textwidth]{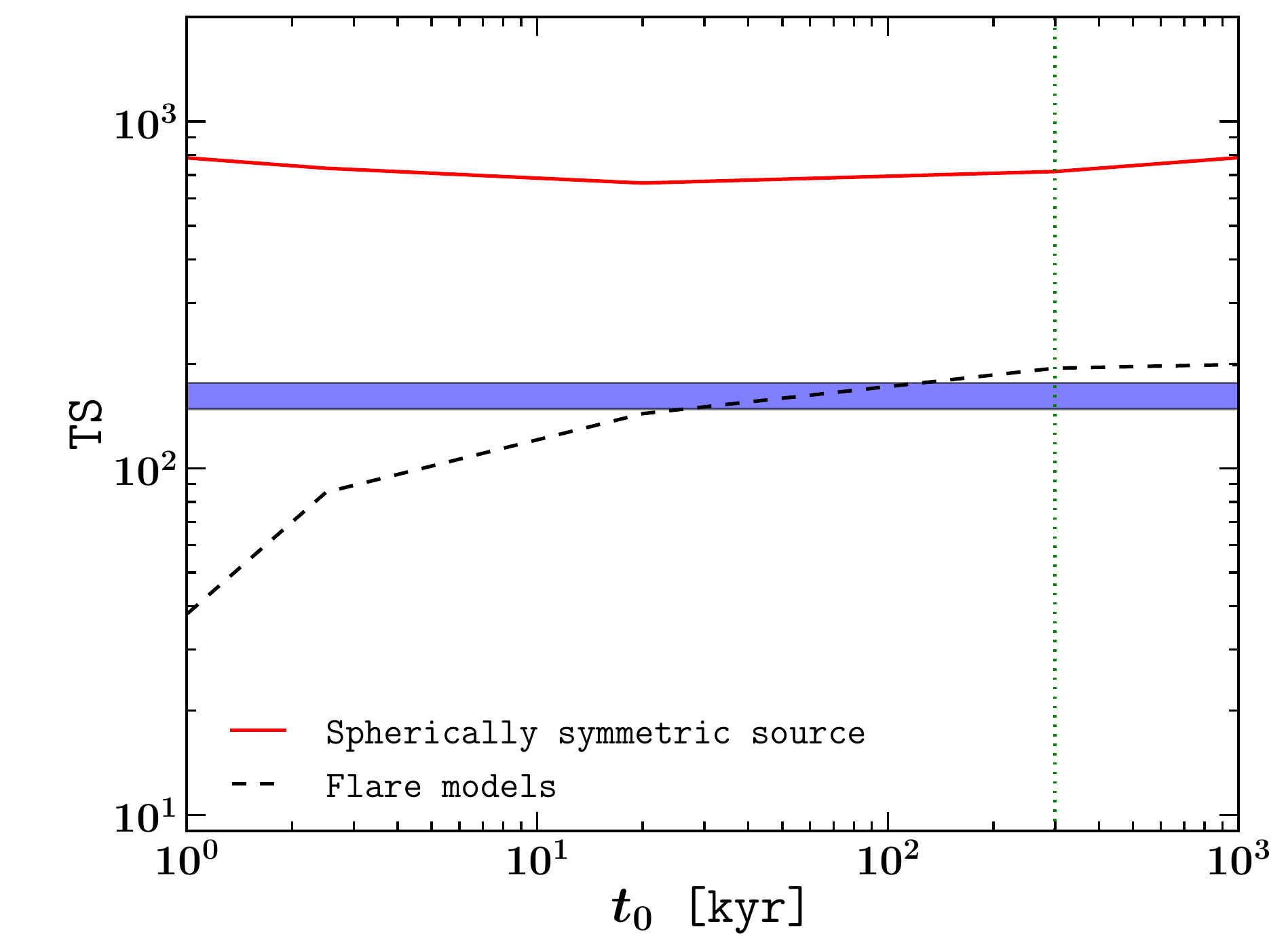}
\end{center}
\caption{ Fermi-tools test statistic (TS) values for the GC spherically symmetric source (red solid line) and several flare models (black dashed line) of different ages ( see also Fig.~\ref{fig:SpatialMorphologyFlareModel}). Note that each flare 
template was included in addition to the spherical source. The blue shaded region encloses the TS values obtained for the Ridge using a 20 cm map or the HESS residuals map~\citep{MaciasGordon2014}. {\color{black}The black dotted line displays the  age needed for the GeV CRs to diffuse to the ridge edges (see Sec.~\ref{sec:nonssfit}).         }}
\label{fig:TSvaluesFlareModels}
\end{figure}

 \begin{figure*}
\begin{center}
\begin{tabular}{cc}

\includegraphics[width=0.5\textwidth]{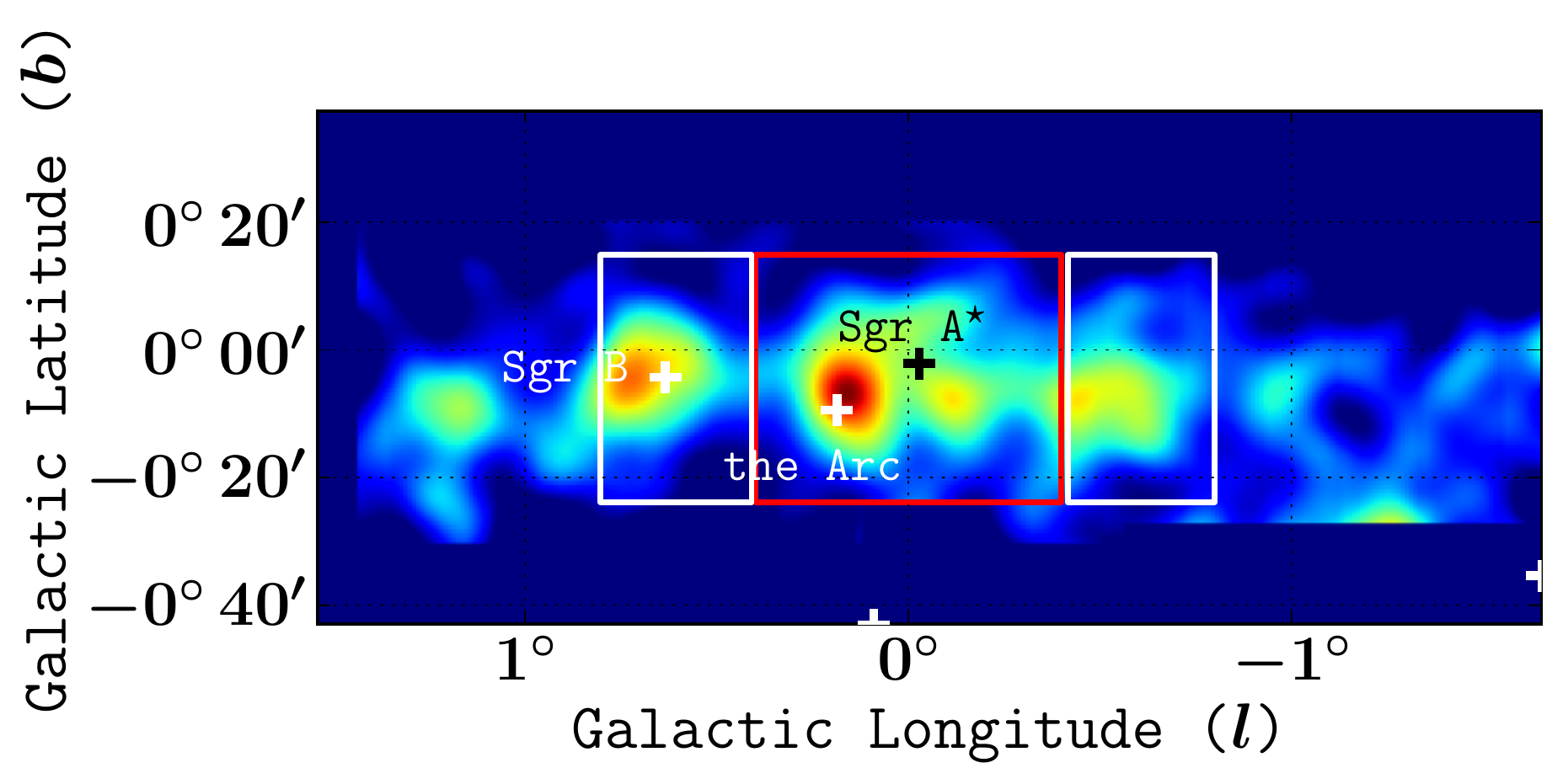} & \includegraphics[width=0.5\textwidth]{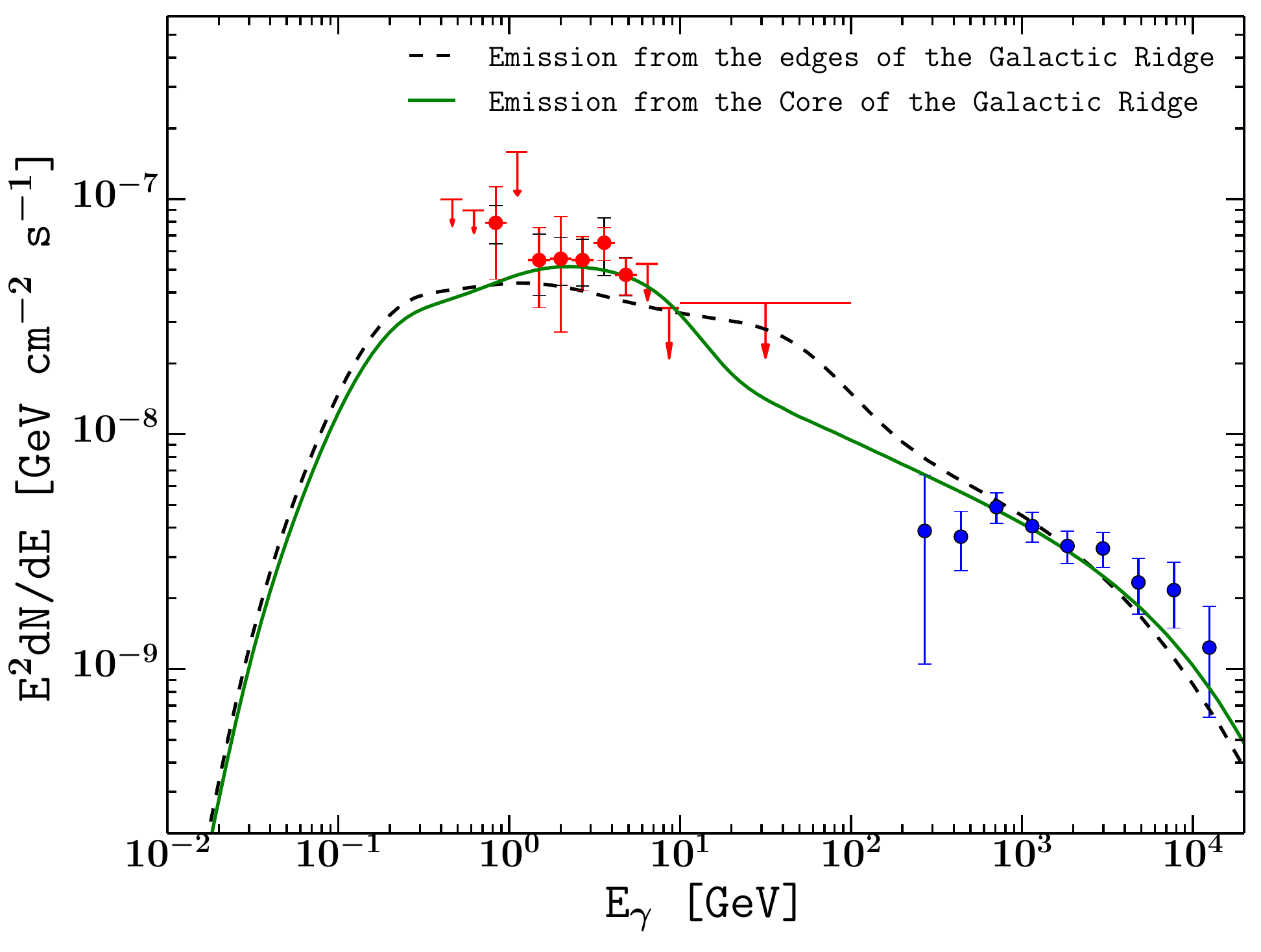}  

\end{tabular}
\end{center}
\caption{ {\bf Spatial variation of the SED along the plane of the Ridge:} \textit{Left:} Background subtracted diffuse TeV-gamma rays image as seen by HESS~\citep{Aharonian:2006au}. The red box encloses the core of the region defined by $-0.4^{\circ}\leq l\leq 0.4^{\circ}$, $\vert b \vert \leq 0.3^{\circ}$ and white boxes the edges $0.4^{\circ} \leq l \leq 0.8^{\circ}$, $\vert b \vert \leq 0.3^{\circ}$ and $-0.8^{\circ} \leq l \leq -0.4^{\circ}$, $\vert b \vert \leq 0.3^{\circ}$. \textit{Right:} Displayed are the predictions of the non-steady-state model (with Kolmogorov diffusion) for the gamma-ray spectra in the regions defined in the left panel. The spectrum at the edges is computed as the mean of the spectrum at both ends.}
\label{fig:spatialindex}

\end{figure*}

{\color{black} 
Altogether, the steady state fitting does not particularly well constrain the ISM and other parameters though we can say with some definiteness that the magnetic field, consistent with previous work \citep{Crocker2010}, has to be at least an order of magnitude larger than in the local ISM.
To go further we  have to impose further, physically-motivated, constraints.
For instance, only allowing the expected $\kappa_{\rm ep}^{\tiny{Bell}}$ motivated by shock acceleration, we are forced to a $\pi^0$-type solution.

Also, although we explored allowing the gas density to float in our steady state fitting, we found that
fit parameters were poorly determined because of degeneracies in this circumstance.
Because of this, we also fix the gas density at the volumetric average value.
In the steady-state model CR particles are removed from the region in an energy-independent fashion consistent with the hard, in situ, steady state, non-thermal particle distribution inferred from the  non-thermal radiation detected from the region.
As we have discussed, this energy-independent escape could, in principle, be affected  by a wind or energy-independent (or weakly dependent) diffusion or streaming.
Solutions with a fast escape and low gas density would then be very plausibly interpreted as a wind, as has been mentioned, given
there is much prior evidence for the existence of a nuclear outflow with about the required speed \citep[cf.][and references therein]{Crocker2012}.}
In our solutions, however, where we have imposed 
 the volumetric average gas density, we have found a 
 short escape time.  
 The only physically plausible interpretation (at least within a single zone scenario) is, then, the energy independent (or very weakly dependent) diffusion or streaming where the cosmic rays 
 escape {\it through} the static gas distribution.
A high gas density ($\gsim$ few $\times 1 $ cm$^{-3}$) together with fast escape {\it in} the gas (advected by the wind) can be ruled out as implying an energetically-implausible mass flux out of the region -- at least in a single zone model; but this may be more revealing of the short-comings of such a treatment than anything else

%
%

%
%
%

Our steady state fitting does not attempt to reproduce the morphology of the non-thermal radiation detected from the region.
Rather it is implicitly posited that this simply reflects the convolution of where the distributed acceleration is taking place throughout the region together with the gas or magnetic field distributions.
Note that, {\it prima facie}, it is not unreasonable that the projected surface density  of particle acceleration sites, presumably related to star formation processes, is correlated with the molecular gas column
{\color{black}(e.g., the largest $H_2$ column is through the giant molecular cloud complex Sagittarius B2 which also hosts the most intense star-formation activity)}.

{\color{black}\subsection{Non-steady-state versus steady-state model}}
In contrast to the steady state, in the non-steady-state scenario,  at a fixed distance and time since a CR injection
event, one expects higher-energy CRs to arrive first so the spectrum is
hardened with respect to the injection distribution.

In the left panel of Fig.~\ref{fig:spatialindex} we divide our region of interest in three sectors: the core is defined by Galactic longitude $-0.4^{\circ}\leq l\leq 0.4^{\circ}$ (red box in Fig.~\ref{fig:spatialindex}) and the edges by $0.4^{\circ} \leq l \leq 0.8^{\circ}$ and $-0.8^{\circ}\leq l \leq -0.4^{\circ}$ (white boxes). We calculate the gamma rays in each of these areas and average the spectrum at the edges (see Fig.~\ref{fig:spatialindex}-(Right)). The spectra heights are normalized to fit the data so as to make their comparison convenient. 
{\color{Black} As can be seen, the current data are not accurate enough to detect the spatial
dependence of the spectrum predicted in the
non-steadystate scenario.}
 Better photon statistics in the $1-100$ GeV range are needed to be able to single out CR diffusion imprints in the gamma-ray data.

We also searched for spectral evidence of diffusion processing in the Fermi-LAT data using the \textsc{Fermi-Tools} analysis software. We created normalized map templates of the regions shown in Fig.~\ref{fig:spatialindex} and modeled the GeV-gamma-ray spectrum with a power-law formula. The spectral index of the diffuse $\sim$GeV emission from the core is $2.33\pm 0.06$ and from the edges $2.43\pm 0.07$ (see Table~\ref{tab:spatialindex}).    Diffusion steepening of the CRs is therefore not currently significant. However, as can be seen from Fig.~\ref{fig:spatialindex}, future  GeV data has the potential to detect this or to rule it out. Such observations could also potentially be used to distinguish between the energy-independent  diffusion scenario proposed by \citet{Aharonian:2006au} and the Kolmogorov diffusion two-flare model we have discussed in this article.
		
\begin{table*}
	
	\begin{tabular}{lcccc}
		\hline\hline
		\centering
		Template & $N_0\;\left[{\rm photons \;MeV^{-1}\;cm^{-2}\;s^{-1}} \right]$  &$\delta N_0\;\left[{\rm photons \;MeV^{-1}\;cm^{-2}\;s^{-1}} \right]$  & $\Gamma$ & $\delta \Gamma$\\  \hline 
		Core&  $3.5\times 10^{-12}$ & $0.5\times 10^{-12}$ & 2.33 &0.06 \\
		Edges& $2.1\times 10^{-12}$ & $0.3\times 10^{-12}$ &2.43 & 0.07\\ \hline\hline 
	\end{tabular}

	\caption{\label{tab:spatialindex} Best fit power-law ($dN/dE=N_{0}(E/E_{0})^{-\Gamma}$) parameters  obtained from a likelihood analysis performed on the regions defined in the left panel of Fig.~\ref{fig:spatialindex}.  Errors are statistical (68\% CI) only and were computed with the \textsc{Fermi-Tools} software. }
\end{table*}

{\color{black}\section{Conclusions}					We have examined steady-state and non-steady-state models of the Ridge gamma-ray excess emission. We showed that a flare model from the GC  provides an acceptable fit to the TeV, GeV and radio data, provided the diffusion coefficient is energy independent. However, if Kolmogorov-type turbulence is assumed to inform the diffusion coefficient, we found that two flares are needed, one for the TeV data (occurring approximately $ 10^4 $ years ago) and an older one for the GeV data (occurring approximately $10^5$ years ago).  

We also found that
the flare models we investigated do not fit the spherically symmetric GeV excess
as well as the usual generalized NFW spatial profile, but are better suited to explaining
the Ridge excess. This is due to the ridge like morphology of  the CMZ gas distribution.

We also showed that, assuming distributed injection, a range of steady-state models are able to explain the GeV, TeV, and radio data for the excess Ridge-like emission but classes of solution
with a floating gas density were poorly constrained.
Fixing the gas density at a high value equal to the volumetric average value we find good solutions that, because of fast escape,  have to be interpreted as energy-independent diffusion or streaming of the escaping cosmic rays {\it through} the gas (rather than escape of particles {\it in} gas advected away on a wind).
Consistent with previous work, we robustly find that the magnetic field had to be at least an order of magnitude larger than in the local ISM for all models considered.

 Additionally, we showed how the flare and steady-state models may be distinguished with future gamma-ray data by looking for a spatial dependence of the gamma-ray spectral index in the GeV range.
}

\section*{Acknowledgments}
OM is supported by a UC Doctoral Scholarship. RMC is the recipient of an Australian Research Council Future Fellowship (FT110100108). SP is supported in part by the US Department of Energy under Contract DE-SC0010107-001. RMC thanks Casey Law for helpful discussions. This work makes use of \textsc{Fermi Science Tools}~\footnote{
}, \textsc{Minuit2}~\citep{minuit} and  \textsc{SciPy}~\citep{scipy}.



\newpage

\bibliography{GalacticRidge} 


\label{lastpage}

\end{document}